\documentclass[journal]{IEEEtran}
\usepackage{cite}
\usepackage{amsmath,amssymb,amsfonts}
\usepackage{graphicx}
\usepackage{textcomp}
\usepackage{wrapfig}
\usepackage[caption=false,font=normalsize,labelfont=sf,textfont=sf]{subfig}
\usepackage{cite}
\usepackage{threeparttable}
\usepackage{booktabs}
\usepackage{multirow}
\usepackage{algorithm}
\usepackage{algorithmicx}
\usepackage{algpseudocode}
\usepackage{siunitx}
\usepackage{amssymb}
\usepackage{bm}
\newcommand{\jj}{\mathrm{j}}
\newcommand{\dd}{\mathrm{d}}

\newcommand{\rr}{\mathrm{r}}
\newcommand{\bb}{\mathrm{b}}
\newcommand{\hh}{\mathrm{h}}

\newcommand{\grm}{\mathrm{g}} 
\newcommand{\Icomp}{\scriptscriptstyle\mathrm{I}}
\newcommand{\Qcomp}{\scriptscriptstyle\mathrm{Q}}
\def\BibTeX{{\rm B\kern-.05em{\sc i\kern-.025em b}\kern-.08em
    T\kern-.1667em\lower.7ex\hbox{E}\kern-.125emX}}
\usepackage{balance}
\begin{document}
\title{Physiology-Informed Multivariate Variational Mode Decomposition for Orientation-Robust Sensing Using Distributed Millimeter-Wave Radar Systems}
\author{Kimitaka~Sumi,~\IEEEmembership{Graduate Student Member,~IEEE},
  Takuya~Sakamoto,~\IEEEmembership{Senior Member,~IEEE}
  \thanks{
  The authors are with the Department of Electrical, Electronic, and Digital Science and Engineering, Graduate School of Engineering, Kyoto University, Kyoto 615-8510, Japan (e-mail: sumi.kimitaka.25r@st.kyotou.ac.jp; sakamoto.takuya.8n@kyoto-u.ac.jp). 
  }}

\maketitle

\begin{abstract}
This study proposes a robust non-contact physiological sensing method using distributed radar systems to address the challenges arising from varying body orientations.
Conventional physiological sensing methods using a single radar system suffer from performance degradation caused by variations in the relative geometry between the radar and the human body, particularly those caused by changes in body orientation.
To help radar systems overcome this limitation, we propose a multivariate physiological variational mode decomposition method that extracts common respiratory and heart rates across multiple displacements of various body parts acquired from distributed radar systems.
The proposed method incorporates harmonic constraints and a gap component tailored to the intrinsic nature of physiological signals into the model, thereby improving both estimation accuracy and robustness.
Simultaneous measurements using four radar systems were performed on six participants under different participant-position and body-orientation conditions.
Using the proposed method, we achieved success rates exceeding 90\% for respiration and heartbeat estimation, which represent improvements of 19.8 and 25.3 percentage points, respectively, over rates using a conventional method based on a single radar system.
Furthermore, we conducted simultaneous measurements of 16 participants using two radar systems in a multi-person scenario and successfully estimated both respiratory and heart rates with over 85\% success rates.
Our work contributes to the realization of practical radar-based non-contact physiological sensing by establishing a data fusion framework for distributed radar systems.
\end{abstract}

\begin{IEEEkeywords}
body orientation, distributed radar systems, millimeter-wave radar, multivariate variational mode decomposition, non-contact physiological sensing, sensor data fusion
\end{IEEEkeywords}

\section{Introduction}
\IEEEPARstart{R}{espiratory} rate and heart rate are important vital signs that can indicate the onset of various diseases \cite{10.1001/jama.1980.03310100041029, 10.1007/BF02600071, 10.1016/S1520-765X(03)90001-0, 10.1016/j.resuscitation.2006.08.020, 0002-838X}. As such, continuous monitoring of these signs is essential. Although contact sensors are commonly used to measure vital signs, they are not suitable for long-term measurements because they can cause issues such as skin irritation, and discomfort.
In contrast, radar, a form of radio wave sensing, has long been investigated for the non-contact measurement of vital signs. 
The advantages of radar include its usability in privacy-sensitive environments and the ability to penetrate clothing and bedding, which facilitates noninvasive and continuous monitoring in daily life.

Despite these advantages, radar-based physiological sensing remains highly sensitive to the relative geometry between the radar and the human body, particularly the distance and body orientation, resulting in degraded estimation accuracy~\cite{10.1109/TMTT.2006.884652, 10.1109/TMTT.2013.2252186, 10.1109/APS.2016.7696293, 10.1145/2971648.2971744, 10.1109/JETCAS.2018.2811339}.
Specifically, previous studies have shown that the accuracy of radar-based sensing decreases when measurements are performed from the side or back of the human body as opposed to from the front.
One possible explanation is that respiratory motion mainly occurs in the dorsoventral direction, and cardiac-induced motion is prominent at the chest wall~\cite{10.1007/BF02441473, 10.1183/09031936.97.10081865, 10.1152/jappl.1997.83.5.1531}; thus, displacement due to respiration and heartbeat can be observed more clearly from the front of the body.
Nevertheless, respiratory motion can also be detected from the lateral side of the body, and cardiac activity can appear as arterial wall motion located in other body regions, such as the neck and calf~\cite{10.1016/j.medengphy.2007.05.008, 10.23919/ISAP47053.2021.9391188, 10.1109/JSEN.2021.3052602}.
These findings suggest that physiological motions are not limited to a localized region but rather are distributed across multiple body parts.
To achieve stable accuracy regardless of body orientation, scientists must  effectively integrate physiological motions observed from these multiple body parts.

Since radar can separate echoes from multiple human body parts through range resolution enabled by wideband transmission and direction-of-arrival (DoA) estimation using antenna arrays, several studies have explored algorithms for selecting or integrating these echoes to improve physiological sensing.
Chang \textit{et al.} and Zhang \textit{et al.} extracted echoes corresponding to individual range bins using frequency-modulated continuous-wave (FMCW) radar~\cite{10.1109/VTC2022-Spring54318.2022.9860799, 10.1145/3589347}.
Wang \textit{et al.} acquired echoes from range-angle maps generated by combining DoA estimation with range separation using an FMCW array radar~\cite{10.1109/JIOT.2021.3128548}.
For echo integration, Chang \textit{et al.} proposed a deep-learning model that estimates weighting parameters for respiratory and heart rates from echoes, and integrated rates by using these estimated weights~\cite{10.1109/VTC2022-Spring54318.2022.9860799}.
Zhang \textit{et al.} selected echoes based on the spectral power of physiological signals and integrated them using cross-correlation techniques~\cite{10.1145/3589347}.
In contrast, Wang \textit{et al.} first separated the respiration and heartbeat components contained in echoes using multivariate variational mode decomposition (MVMD) and then integrated them by summing the corresponding components across the echoes~\cite{10.1109/JIOT.2021.3128548}.
However, these conventional studies all use single-radar systems and are therefore limited to reflected regions of the human body surface from a single viewpoint, which makes it difficult to achieve robust performance for arbitrary body orientations.
This limitation becomes even more pronounced in multi-person scenarios, where intersubject blocking weakens or completely occludes the echoes from a target individual, thereby making estimation more challenging \cite{10.1088/1361-6560/ae22b8}.

To address this limitation, systems employing multiple radar antennas positioned around the human body have been proposed.
Walterscheid and Smith experimentally demonstrated that the signal-to-noise ratio of physiological signals can be improved by superimposing the velocity signals estimated from a 4-transmit 4-receive ultrawideband (UWB) radar system~\cite{10.1109/EMBC.2017.8037598}.
Iwata \textit{et al.} proposed an integration method robust to human-body shadowing in multi-person scenarios by selecting echoes based on the kurtosis of the respiratory component using two FMCW radar systems~\cite{10.1109/JSEN.2021.3117707}.
Ren \textit{et al.} designed a 1-transmit 4-receive stepped-frequency continuous-wave (SFCW) radar system and directly estimated respiratory and heartbeat components by applying derivative independent component analysis to displacements acquired from different viewpoints~\cite{10.1109/TMTT.2021.3101655}.
Other studies have used multiple radar systems mainly to cancel body movement artifacts but have not addressed arbitrary orientations \cite{10.1109/TMTT.2008.2007139, 10.1109/TAES.2020.2990817, 10.1109/AP-S/INC-USNC-URSI52054.2024.10686644}. Although these multiradar systems have been investigated, multivariate signal processing methods to extract commonly shared physiological signals from diverse displacements of multiple body parts via multiple-viewpoint antennas have yet to be established.

In this study, we propose a multivariate physiological mode variational decomposition (MPVMD) method to achieve robust physiological signal sensing with distributed radar systems.
The proposed method involves formulating a variational optimization problem, in which the center frequencies of respiration and heartbeat are modeled as common parameters shared across multiple displacements acquired from different radar viewpoints, with constraints imposed on the harmonic components.
This formulation enables the direct estimation of the frequencies by jointly exploiting the harmonic information distributed across all signals.
Furthermore, we propose a gap component in the model to improve robustness against unwanted components located between the respiration and heartbeat frequency bands.
The proposed algorithm is mathematically built upon and extends the conventional variational mode decomposition (VMD) technique~\cite{10.1109/TSP.2013.2288675}, which has demonstrated effectiveness in noise suppression and physiological signal separation for radar-based physiological sensing~\cite{10.23919/Eusipco47968.2020.9287524, 10.1109/JIOT.2021.3075167, 10.1109/INFOCOM48880.2022.9796912, 10.1109/ACCESS.2025.3575932, 10.1109/JSEN.2024.3494755}.
Consequently, the proposed method can determine physiological properties while inheriting the advantages of VMD, enabling the stable and efficient extraction of physiological signals from distributed radar systems.

In contrast to recent deep learning approaches \cite{10.1109/TIM.2023.3300471, 10.1109/ACCESS.2024.3482690, 10.1109/JSEN.2025.3646149}, our method applies an interpretable framework based on a mathematical model that leverages the harmonic structures of quasiperiodic physiological signals.
This design choice is motivated by the fact that no large-scale annotated datasets for distributed radar-based physiological measurements are publicly available, as well as the fact that the algorithms used in deep learning models may not always be straightforward to interpret.
A preprint of this manuscript has been posted online \cite{arXiv:2510.10542}. The main contributions of this work are as follows.
\begin{itemize}
\item We propose MPVMD as a multivariate signal processing method for robust, non-contact physiological sensing using distributed radar systems.
\item We conducted experiments using four radar systems under different participant positions and body-orientation conditions. Compared with the conventional method that involves applying VMD~\cite{10.1109/TSP.2013.2288675} to a velocity signal obtained from a single-radar system, the proposed method demonstrates superior estimation accuracy for both respiration and heartbeat center frequencies.
\item We also compared the proposed method with MVMD \cite{10.1109/TSP.2019.2951223} as a conventional multivariate signal processing method. To the best of our knowledge, this is the first evaluation of MVMD for distributed radar-based physiological sensing, and the results demonstrate that the proposed method provides superior heartbeat estimation performance compared with MVMD.
\item We further conducted simultaneous measurements of 16 participants using two radar systems in a multi-person scenario with closely spaced participants and successfully estimated both respiratory and heart rates with over 85\% success rates.
\end{itemize}

\section{Preliminaries}
This study proposes the MPVMD method based on VMD. 
For clarity, we first describe the principle of microdisplacement measurement using FMCW radar.
We then outline how conventional univariate VMD is used to extract respiration and heartbeat data from the displacement data.
Finally, we introduce its multivariate extension, i.e., MVMD, which enables the joint analysis of multiple displacements and acts as the mathematical foundation for our proposed MPVMD.

\subsection{Microdisplacement Measurement Using Frequency-Modulated Continuous-Wave Radar}\label{sec:C2Radar}
FMCW radar is capable of measuring target ranges.
The transmitted signal of the FMCW radar is given as follows:
\begin{equation}
  s^{\mathrm{T}}(\tau) = A\exp{\left\{\jj(2\pi f_0 \tau + \pi \gamma \tau^2)\right\}}, 
\end{equation}
where $A$ is the amplitude of the transmitted signal, $f_0$ is the starting frequency of the transmission, and $\tau$ denotes the fast time. The chirp rate is
defined as $\gamma=B/T_{\mathrm{c}}$, where $B$ is the sweep bandwidth and $T_{\mathrm{c}}$ is the sweep duration.

Let $M$ denote the number of elements in the 1-D virtual array, and let $s_m(\tau,t)$ be the intermediate-frequency (IF) signal, which is obtained by mixing the transmitted and received signals, where $m$ is the index of elements in the array and $t$ is the slow time.
In this study, a complex radar image is calculated by applying a range Fourier transform and beamforming to the received IF signals, which is expressed mathematically as follows:
\begin{align}
I(r,\theta,t) = \sum_{m=1}^{M} w_m^{\mathrm{A}} a^*_m(\theta) s'_m(r,t),
\end{align}
where $r$ is the range, $\theta$ is the azimuth, $w_m^{\mathrm{A}}$ is the antenna weighting (Taylor window), $a^*_m(\theta)$ is the conjugated steering vector, and $s'_m(r,t)$ is the range-processed signal from the $m$-th antenna, defined as follows:
\begin{align}
  s'_{m}(r,t) = \int_{0}^{T_{\mathrm{c}}} w^{\mathrm{R}}(\tau) s_{m}(\tau,t)\exp\!\left(-\jj2\pi \tfrac{2Br}{cT_{\mathrm{c}}}\tau\right) \dd\tau,
\end{align}
where $w^{\mathrm{R}}(\tau)$ is the range window function and $c$ is the speed of light.

Consider an ideal point target located at $(r^{*}, \theta^{*})$. Assuming that the target undergoes a microdisplacement $d(t)$, the distance between the target and the antenna is represented by $R(t)=r^{*}+d(t)$.
The target displacement is observed as phase modulation of the complex signal $I(r^{*}, \theta^{*}, t)$ as follows:
\begin{align}
I(r^{*}, \theta^{*},t) \propto \exp\left(  \jj \frac{4\pi f_0 d(t)}{c} \right).
\end{align}
The microdisplacement of a human body surface can thus be estimated by demodulating the complex signal. 
However, the estimated displacement is a superposition of respiration, heartbeat, and unwanted random body movements.
Therefore, these components must be effectively decomposed to achieve reliable physiological signal estimation using techniques such as VMD, as explained in the next section.

\subsection{Variational Mode Decomposition}\label{sec:C2VMD}
VMD is one of the time-frequency analysis techniques that adaptively decomposes an input signal into multiple narrowband signals, referred to as intrinsic mode functions (IMFs)~\cite{10.1109/TSP.2013.2288675}.
Let $x(t)$ be a real-valued signal such as the microdisplacement of a human body surface, which corresponds to the time derivative of the displacement $d(t)$ in this study. The signal $x(t)$ is decomposed into its constituent IMFs, formulated as:
\begin{align}
  x(t) &= \sum_{k=1}^{K} u_k(t), \\
  u_k(t) &= a_k(t) \cos(\phi_k(t)),
\end{align}
where $u_k(t)$ denotes the $k$th IMF, $K$ is the total number of IMFs, $a_k(t)$ is the envelope, and $\phi_k(t)$ is the instantaneous phase. Each IMF is assumed to be narrowband, which means that the variation of $a_k(t)$ is much slower than that of $\phi_k(t)$.

The decomposition into IMFs is formulated as the following optimization problem:
\begin{align}
  \begin{split}
    \underset{\{ u_{k}(t), \omega_{k} \}_{k=1}^{K}} {\operatorname{minimize}} \: 
    & \sum^{K}_{k=1}
    \left\|\frac{\partial}{\partial t} \left[ u^{+}_{k}(t) \mathrm{e}^{-\jj\omega_{k}t} \right] \right\|^{2}  \text{s.t.} \, \sum^{K}_{k=1} u_{k}(t) = x(t),
  \end{split}
  \label{eq:opt_VMD}
\end{align}
where $\omega_k$ is the center frequency of the $k$th IMF. Here, $u^{+}_k(t)$ is the analytic signal of $u_k(t)$, defined using the Hilbert transform $\mathcal{H}[\cdot]$ as
\begin{align}
  u^{+}_{k}(t) &= u_{k}(t) + \jj \mathcal{H}[u_{k}(t)], \\
  \mathcal{H}[u_{k}(t)] &= \frac{1}{\pi} \int_{-\infty}^{\infty} \frac{u_{k}(\tau)}{t-\tau} \, \dd \tau.
\end{align}

By introducing a Lagrange multiplier $\lambda(t)$, the objective function in Eq.~\eqref{eq:opt_VMD} can be expressed as the following unconstrained optimization problem:
\begin{align}
  \mathcal{L}(\{ u_{k}(t), & \omega_{k} \},\lambda(t)) = \alpha \sum^{K}_{k=1} \left\|\frac{\partial}{\partial t} \left[ \mathrm{e}^{-\jj\omega_{k}t} u^{+}_{k}(t) \right] \right\|^{2}  \nonumber \\
& +\left\| x(t) - \sum^{K}_{k=1} u_{k}(t) \right\|^{2}+\left\langle \lambda(t), x(t) - \sum^{K}_{k=1} u_{k}(t) \right\rangle
\end{align}
where $\alpha$ is the hyperparameter that controls the narrowband characteristics of the IMFs and $\langle \cdot, \cdot \rangle$ denotes the inner product of two functions.

An equivalent frequency-domain formulation of the objective function is obtained by applying Parseval's theorem:
\begin{align}
    \hat{\mathcal{L}}(\{ \hat{u}_{k}(\omega), \omega_{k}\}, \hat{\lambda}(\omega))&=
    4\alpha \sum^{K}_{k=1}  \int_0^{\infty} (\omega-\omega_k)^2|\hat{u}_{k}(\omega)|^2\dd\omega  \nonumber\\
                                                           &+\left\| \hat{x}(\omega) - \sum^{K}_{k=1} \hat{u}_{k}(\omega) + \frac{\hat{\lambda}(\omega)}{2} \right\|^{2}                      
\end{align}
where $\hat{x}(\omega)$,$\hat{u}_{k}(\omega)$, and $\hat{\lambda}(\omega)$ denote the Fourier transforms of $x(t)$,$u_{k}(t)$, and $\lambda(t)$, respectively.

This optimization problem is solved using the alternating direction method of multipliers (ADMM)~\cite{10.1561/2200000016}. In the ADMM framework, the parameters $\{ \hat{u}_{k}(\omega)\}$, $\{\omega_{k}\}$, and $\hat{\lambda}(\omega)$ are iteratively updated by alternately solving their respective subproblems. The update rule for each parameter is given by:
\begin{align}
  \hat{u}_{k}^{n+1}(\omega) & = \frac{\hat{x}(\omega) - \sum_{k'\neq k} \hat{u}_{k'}^{n}(\omega) + \hat{\lambda}^{n}(\omega)/2}{1 + 2\alpha(\omega - \omega_k^{n})^2},   \\
  \omega_k^{n+1}  & =  \frac{\int_0^{\infty} \omega |\hat{u}_{k}^{n+1}(\omega)|^2 \, \dd \omega}{\int_0^{\infty} |\hat{u}_{k}^{n+1}(\omega)|^2 \, \dd \omega}, \\
  \hat{\lambda}^{n+1}(\omega)          & = \hat{\lambda}^{n}(\omega) + \eta \left( \hat{x}(\omega) - \sum_{k=1}^{K} \hat{u}_{k}^{n+1}(\omega) \right),
\end{align}
where $n$ is the iteration index of the ADMM algorithm, $\sum_{k'\neq k}$ denotes summation over all modes $k'$ except for $k$, and $\eta$ represents the step size for the Lagrange multiplier.

The objective function in Eq.~\eqref{eq:opt_VMD} penalizes the temporal bandwidth of each analytic signal $u^{+}_k(t)$ after demodulation by $\omega_k$, thus enforcing the narrowband property of the extracted IMFs. As a result, $x(t)$ is decomposed into a set of narrowband components whose amplitudes and phases vary over time.

\subsection{Multivariate Variational Mode Decomposition}
MVMD is an extension of VMD for multichannel signals, designed to extract IMFs that share common center frequencies across channels \cite{10.1109/TSP.2019.2951223}. 

Given a multivariate real signal 
\begin{align}
  \bm{x}(t) = [x_1(t), x_2(t), \ldots, x_C(t)]^{\top},
\end{align}
where $C$ is the number of channels, MVMD decomposes it into narrowband components $u_{k,c}(t)$ for mode $k$ and channel $c$. Each multivariate IMF is expressed as
\begin{align}
  \bm{u}_k(t) = [u_{k,1}(t), \ldots, u_{k,C}(t)]^{\top}.
\end{align}

The optimization problem is formulated as
\begin{align}
  \underset{\{ \bm{u}_{k}(t), \omega_{k}\}_{k=1}^{K}}{\operatorname{minimize}} 
   \: \sum^{K}_{k=1}
 \left\|\frac{\partial}{\partial t} \left[ \mathrm{e}^{-\jj\omega_{k}t} \bm{u}^{+}_{k}(t) \right] \right\|^{2} \text{s.t.}\: \sum^{K}_{k=1} \bm{u}_{k}(t) = \bm{x}(t),
  \label{eq:opt_multiVMD}
\end{align}
where $\omega_k$ is the common center frequency of the $k$th IMF across all channels and $ \bm{u}^{+}_k(t) = [u^{+}_{k,1}(t),\ldots, u^{+}_{k,C}(t)]^{\top}$ denotes the analytic signal of $\bm{u}_{k}(t)$.

Similar to the VMD algorithm in Section~\ref{sec:C2VMD}, the optimization problem in Eq.~\eqref{eq:opt_multiVMD} is solved using ADMM, where the update rule for each parameter subproblem is given by:
\begin{align}
  \hat{u}_{k,c}^{n+1}(\omega) & = \frac{\hat{x}_c(\omega) - \sum_{k'\neq k} \hat{u}_{k',c}^{n}(\omega) + \hat{\lambda}_c^{n}(\omega)/2}{1 + 2\alpha(\omega - \omega_k^{n})^2},         \\
  \omega_k^{n+1}              & =  \frac{\sum_{c=1}^{C}\int_0^{\infty} \omega |\hat{u}_{k,c}^{n+1}(\omega)|^2 \, \dd \omega}{\sum_{c=1}^{C}\int_0^{\infty} |\hat{u}_{k,c}^{n+1}(\omega)|^2 \, \dd \omega}, \\
  \hat{\lambda}_c^{n+1}(\omega)          & = \hat{\lambda}_c^{n}(\omega) + \eta \left( \hat{x}_c(\omega) - \sum_{k=1}^{K} \hat{u}_{k,c}^{n+1}(\omega) \right),
\end{align}
where $\hat{x}_c(\omega)$, $\hat{u}_{k, c}(\omega)$, and $\hat{\lambda}_c(\omega)$ are the Fourier transforms of $x_c(t)$, $u_{k,c}(t)$, and $\lambda_c(t)$, respectively.

If conventional univariate VMD is applied independently to each channel, the center frequencies are estimated separately for each signal. 
In contrast, MVMD enforces consistency by extracting IMFs with shared modes $\omega_k$ across all channels. 
However, MVMD does not explicitly model the constraints on harmonic characteristics caused by the quasiperiodicity of physiological signals; therefore, the proposed method incorporates these properties.

\section{Proposed Multivariate Physiological Variational Mode Decomposition Method}
\subsection{Proposed Model}
The proposed method decomposes the input signal $\bm{x}(t)$ into respiration, heartbeat, trend, and gap components, which are modeled as follows:
\begin{align}
  x_c(t)                  & = x_{\mathrm{model}, c}(t) + x_{\mathrm{res}, c}(t),                                              \label{eq:propModelTimeDomainRes} \\
  x_{\mathrm{model}, c}(t) & = \sum_{i=1}^{K_\mathrm{r}} r_{c,i}(t) + \sum_{j=1}^{K_\mathrm{h}} h_{c,j}(t) + b_c(t) + g_c(t), \label{eq:propModelTimeDomain}
\end{align}
where $x_{\mathrm{res}, c}(t)$ is the residual component, $r_{c, i}(t)$ is the $i$-th harmonic component of respiration, $h_{c,j}(t)$ is the $j$-th harmonic component of the heartbeat, and $K_\mathrm{r}$ and $K_\mathrm{h}$ are the total number of harmonics for the respiration and heartbeat signals, respectively. 
In addition, $b_c(t)$ is the trend, which captures frequency components even lower than those of respiration, and 
$g_c(t)$ is the gap component, which models the frequency band between the respiration and heartbeat bands.
Specifically, let $\mathcal{E}_{\mathrm{RR}} = [f^{\mathrm{RR}}_{\mathrm{L}}, f^{\mathrm{RR}}_{\mathrm{U}}]$ and $\mathcal{E}_{\mathrm{HR}} = [f^{\mathrm{HR}}_{\mathrm{L}}, f^{\mathrm{HR}}_{\mathrm{U}}]$ denote the frequency bands containing the fundamental frequencies of respiration and heartbeat, where  $f^{(\cdot)}_{\mathrm{L}}$ and $f^{(\cdot)}_{\mathrm{U}}$ represent the respective lower and upper bounds. Based on typical physiological knowledge~\cite{10.1016/0002-9149(92)90947-W, https://www.ncbi.nlm.nih.gov/sites/books/NBK537306/}, we assume the relationship $f^{\mathrm{RR}}_{\mathrm{U}}~<~f^{\mathrm{HR}}_{\mathrm{L}}$. Consequently, the gap frequency band is defined as $\mathcal{E}_{\mathrm{g}}=[f^{\mathrm{RR}}_{\mathrm{U}}, f^{\mathrm{HR}}_{\mathrm{L}}]$. In this study, we set $\mathcal{E}_{\mathrm{RR}}=[0.08, \,0.4]$~Hz and $\mathcal{E}_{\mathrm{HR}}=[0.8,\,1.7]$~Hz based on prior knowledge of respiratory and heartbeat rates. This gap component is introduced to improve robustness against unwanted nonphysiological components that may exist between the respiration and heartbeat frequency bands.

The optimization problem of the proposed method is formulated by assuming that the residual component in Eq.~\eqref{eq:propModelTimeDomainRes} is negligible, as follows:
\begin{align}
   & \underset{( \bm{r}_{\mathrm{all}}(t), \bm{h}_{\mathrm{all}}(t), \bm{b}(t), \bm{g}(t), \omega_{\rr}, \omega_{\hh} ) }  {\operatorname{minimize}} \quad J_{\rr} + J_{\hh} +  J_{\bb} +  J_{\grm} \nonumber \\
   & \text{s.t.} \quad \bm{x}(t) = \sum_{i=1}^{K_\mathrm{r}} \bm{r}_{i}(t) + \sum_{j=1}^{K_\mathrm{h}} \bm{h}_{j}(t) + \bm{b}(t) + \bm{g}(t),
  \label{eq:opt_MPVMD}
\end{align}
where $\bm{r}_i(t), \bm{h}_j(t), \bm{b}(t)$, and $\bm{g}(t)$ are the $C$-channel multivariate respiration, heartbeat, trend, and gap signals,  respectively. We define $\bm{r}_{\mathrm{all}}(t) = [ \bm{r}_1(t)^{\top}, \ldots, \bm{r}_{K_\mathrm{r}}(t)^{\top} ]^{\top}$ to be the aggregate respiration signal including all harmonics, while $\bm{h}_{\mathrm{all}}(t) = [ \bm{h}_1(t)^{\top}, \ldots, \bm{h}_{K_\mathrm{h}}(t)^{\top} ] ^{\top}$ is the aggregate heartbeat signal including all harmonics. The parameters $\omega_{\mathrm{r}}$ and $\omega_{\mathrm{h}}$ indicate the center frequencies of respiration and heartbeat, respectively, and 
$J_\mathrm{r}$, $J_\mathrm{h}$, $J_\mathrm{b}$, and $J_\mathrm{g}$ are the bandwidth-related cost functions of each respective component, defined as follows:
\begin{align}
  J_\rr                 & = \alpha_{\rr}  \sum_{c=1}^{C} \sum_{i=1}^{K_\mathrm{r}} \left\| \frac{\partial}{ \partial t} \left[r^{+}_{c,i}(t)\exp\{-\jj(i\omega_{\rr}t)\}\right] \right\|^2, \label{eq:Jr} \\
  J_\hh                 & = \alpha_{\hh}  \sum_{c=1}^{C} \sum_{j=1}^{K_\mathrm{h}} \left\| \frac{\partial}{ \partial t} \left[h^{+}_{c,j}(t)\exp\{-\jj(j\omega_{\hh}t)\}\right] \right\|^2, \label{eq:Jh} \\
  J_\bb                 & = \alpha_{\bb}  \sum_{c=1}^{C} \left\| \frac{\partial}{ \partial t} b^{+}_{c}(t) \right\|^2,                                                                \label{eq:Jb}      \\
  J_\grm                & = \alpha_{\grm} \sum_{c=1}^{C} \left\| \frac{\partial}{ \partial t} \left[g^{+}_{c}(t)\exp\{-\jj\omega_{\mathrm{g}}t\}\right] \right\|^2   \nonumber \\
                        & + \sum_{c=1}^{C} \sum_{i=1}^{K_\mathrm{r}} \left\| \beta_i(t)*g_c(t) \right\|^2,  \label{eq:Jg} 
\end{align}
where $\alpha_{\mathrm{r}}, \alpha_{\mathrm{h}}, \alpha_{\mathrm{b}}$, and $\alpha_{\mathrm{g}}$ are hyperparameters that control the narrowband characteristics of the corresponding components. Additionally, $\omega_{\grm}$ is the center frequency of the gap component, and 
$\beta_i(t)$ is a filter with a peak at the $i$th harmonic frequency of respiration, so that the gap component is updated to avoid these frequencies.
In the frequency domain, the filter is defined as $\hat{\beta}_i(\omega)=1/\alpha_{\mathrm{r}}(|\omega|-i\omega_{\mathrm{r}})^2$.
Eq.~\eqref{eq:Jb} fixes the center frequency of the trend component to be zero, thereby enabling the representation of a low-frequency trend. For the gap component in Eq.~\eqref{eq:Jg}, the center frequency $\omega_{\grm}$ is fixed to be  $\omega_{\grm}/2\pi=(f^{\mathrm{RR}}_{\mathrm{U}}+f^{\mathrm{HR}}_{\mathrm{L}})/2$. In this study, $\omega_{\grm}/2\pi$ is set to 0.6 Hz.

It should be noted from Eq.~\eqref{eq:Jr} that the harmonic signals $r_{c,i}(t)$ for $i=1,\ldots,K_\rr$ are constrained by a common fundamental frequency $\omega_\rr$. Similarly, in Eq.~\eqref{eq:Jh}, the heartbeat harmonic signals share the common fundamental frequency $\omega_\hh$. By jointly exploiting information from these harmonics, the proposed method can estimate the respiratory and heartbeat center frequencies more robustly.
Furthermore, conventional MVMD in Eq.~\eqref{eq:opt_multiVMD} inherently requires a post-hoc selection of respiratory and heartbeat modes from $K$ decomposed modes. By contrast, we directly model only $\omega_\rr$ and $\omega_\hh$ as the parameters to be estimated in the proposed method. Consequently, it does not require an additional mode-selection procedure and can be used to directly obtain the center frequencies and modes corresponding to respiration and heartbeat.

Similar to conventional VMD, the objective function in the time domain given by Eq.~\eqref{eq:opt_MPVMD} can be reformulated in the frequency domain as $\hat{\mathcal{L}}( \hat{\bm{r}}_{\mathrm{all}}(\omega), \hat{\bm{h}}_{\mathrm{all}}(\omega), \hat{\bm{b}}(\omega), \hat{\bm{g}}(\omega), \omega_{\rr}, \omega_{\hh} )$ expressed as follows:
\begin{align}
  \hat{\mathcal{L}} & =  \hat{J}_{\rr} +  \hat{J}_{\hh} + \hat{J}_{\bb} +  \hat{J}_{\grm} \nonumber \\
& + \sum_{c}^{C} \left\| \hat{x}_c(\omega) - \hat{x}_{\mathrm{model}, c}(\omega) + \frac{\hat{\lambda}_c(\omega)}{2} \right\|^2, \label{eq:objFuncMPVMDFreq} \\
  \hat{x}_{\mathrm{model}, c}(\omega) & = \sum_{i=1}^{K_\mathrm{r}} \hat{r}_{c,i}(\omega) + \sum_{j=1}^{K_\mathrm{h}} \hat{h}_{c,j}(\omega) + \hat{b}_c(\omega) + \hat{g}_c(\omega), \label{eq:modelPropFreq}
\end{align}
\begin{align}
  \hat{J}_\rr  & = 4 \alpha_{\rr} \sum_{c=1}^{C} \sum_{i=1}^{K_\mathrm{r}} \int_{0}^{\infty} (\omega - i\omega_\rr)^2 |\hat{r}_{c,i}(\omega)|^2 \dd\omega,                   \\
  \hat{J}_\hh  & = 4 \alpha_{\hh} \sum_{c=1}^{C} \sum_{j=1}^{K_\mathrm{h}} \int_{0}^{\infty} (\omega - j\omega_\hh)^2 |\hat{h}_{c,j}(\omega)|^2 \dd\omega,                   \\
  \hat{J}_\bb  & = 4 \alpha_{\bb} \sum_{c=1}^{C} \int_{0}^{\infty} \omega^2 |\hat{b}_{c}(\omega)|^2 \dd\omega,                                                                \\
  \hat{J}_\grm & = 4 \alpha_{\grm} \sum_{c=1}^{C} \int_{0}^{\infty} (\omega - \omega_{\mathrm{g}})^2 |\hat{g}_{c}(\omega)|^2 \dd\omega                                        \\
               & + 2 \sum_{c=1}^{C} \sum_{i=1}^{K_\mathrm{r}} \int_{0}^{\infty} \frac{1}{\alpha_\rr^2(|\omega|-i\omega_{\rr})^4} |\hat{g}_{c}(\omega)|^2 \dd\omega,
\end{align}
where $\hat{J}_\rr$, $\hat{J}_\hh$, $\hat{J}_\bb$, and $\hat{J}_\grm$ are the frequency-domain representations of the corresponding time-domain cost functions; $\hat{r}_{c,i}(\omega)$, $\hat{h}_{c,j}(\omega)$, $\hat{b}_{c}(\omega)$, and $\hat{g}_{c}(\omega)$ are the Fourier transforms of $r_{c,i}(t)$, $h_{c,j}(t)$, $b_{c}(t)$, and $g_{c}(t)$, respectively; and $\hat{x}_{\mathrm{model}, c}(\omega)$ is the Fourier transform of the model signal for the $c$th channel.

\subsection{Update Rules}
The optimization problem \eqref{eq:objFuncMPVMDFreq} is solved using ADMM. In each step, only one of the parameters from the tuple $(\hat{\bm{r}}_{\mathrm{all}}(\omega), \hat{\bm{h}}_{\mathrm{all}}(\omega), \hat{\bm{b}}(\omega), \hat{\bm{g}}(\omega), \omega_{\rr}, \omega_{\hh})$ is updated while the remaining parameters are fixed. The updates are performed alternately for all parameters and repeated iteratively until convergence is achieved.
The overall procedure of the proposed method is summarized in Algorithm~\ref{alg:PropMPVMD}.
\begin{algorithm}[tb]
  \caption{Proposed multivariate physiological variational mode decomposition algorithm}
  \label{alg:PropMPVMD}
  \addtolength{\baselineskip}{2pt}
  \begin{algorithmic}[]
    \State Initialize
    \State $\hat{r}_{c,i}^1(\omega), \hat{h}_{c,j}^1(\omega), \hat{b}_{c}^1(\omega), \hat{g}_{c}^1(\omega), \hat{\lambda}_{c}^1(\omega),n\gets 0$
    \State $\omega^1_{\rr} \gets \omega_{\mathrm{iniRR}},  \omega^1_\hh \gets  \omega_{\mathrm{iniHR}}$
    \Repeat
      \State  $n \gets n+1$
      \For{$c=1, 2, \ldots, C$}
      \State 1) Update mode $\hat{r}_{c, i}(\omega)$
        \For{$i=1, 2, \ldots, K_\rr$}
        \State {\small $\hat{x}^n_{\rr, c, i}(\omega)=\sum_{i'<i}\hat{r}_{c, i'}^{n+1}(\omega)+\sum_{i'>i}\hat{r}_{c, i'}^{n}(\omega)$}
        \Statex {\small\hspace{9em}$+\sum_{j=1}^{K_\mathrm{h}}\hat{h}_{c,j}^n(\omega)+\hat{b}_c^n(\omega)+\hat{g}_c^n(\omega)$}
        \Statex $\hspace{4em}\hat{r}_{c,i}^{n+1}(\omega) \gets \frac{\hat{x}_c(\omega) - \hat{x}^n_{\rr, c, i}(\omega) + \hat{\lambda}^n_c(\omega)/2}{1 + 2\alpha_\rr (\omega-i\omega_\rr^n)^2}$
        \EndFor
      \State 2) Update mode $\hat{h}_{c, j}(\omega)$
        \For{$j=1, 2, \ldots, K_\hh$} 
        \State {\small $\hat{x}^n_{\hh, c, j}(\omega)=\sum_{j'<j}\hat{h}_{c, j'}^{n+1}(\omega)+\sum_{j'>j}\hat{h}_{c, j'}^{n}(\omega)$}
        \Statex {\small\hspace{9em}$+\sum_{i=1}^{K_\rr}\hat{r}_{c, i}^{n+1}(\omega)+\hat{b}_c^n(\omega)+\hat{g}_c^n(\omega)$}
        \Statex $\hspace{4em}\hat{h}_{c,j}^{n+1}(\omega) \gets \frac{\hat{x}_c(\omega) - \hat{x}^n_{\hh, c, j}(\omega) + \hat{\lambda}^n_c(\omega)/2}{1 + 2\alpha_\hh (\omega-j\omega_\hh^n)^2}$
        \EndFor
        \State 3) Update mode $\hat{b}_c(\omega)$:
        \State \hspace{0em}{\small $\hat{x}^n_{\bb, c}(\omega)=\sum_{i=1}^{K_\rr}\hat{r}_{c, i}^{n+1}(\omega)+ \sum_{j=1}^{K_\hh}\hat{h}_{c, j}^{n+1}(\omega)+\hat{g}_c^n(\omega)$}
        \Statex \hspace{0em}$\hspace{3em}\hat{b}_{c}^{n+1}(\omega) \gets \frac{\hat{x}_c(\omega) - \hat{x}^n_{\bb, c}(\omega) + \hat{\lambda}^n_c(\omega)/2}{1 + 2\alpha_\bb\omega^2}$
        \State 4) Update mode $\hat{g}_c(\omega)$:
        \State\hspace{0em}{\small $\hat{x}^{n}_{\grm, c}(\omega)\!=\!\sum_{i=1}^{K_\rr}\hat{r}_{c, i}^{n+1}(\omega)\!+\!\sum_{j=1}^{K_\hh}\hat{h}_{c, j}^{n+1}(\omega)\!+\!\hat{b}_c^{n+1}(\omega)$}
        \Statex \hspace{0em}$\hspace{3em}\hat{g}_{c}^{n+1}(\omega) \gets \frac{\hat{x}_c(\omega) - \hat{x}^n_{\grm, c}(\omega) + \hat{\lambda}^n_c(\omega)/2}{1 + 2\alpha_\grm(\omega-\omega_\grm)^2 + \sum_{i=1}^{K_\rr} 1/\{\alpha_\rr^2(\omega-i\omega_\rr^n)^4\} }$
      \EndFor
        \State 5) Update center frequency $\omega_\rr$ and $\omega_\hh$ 
        \State \hspace{1em}$\omega_{\text{r}}^{n+1} \gets \frac{\sum_{c=1}^{C}\sum_{i=1}^{K_{\text{r}}} i \int_0^\infty \omega |\hat{r}^{n+1}_{i,c}(\omega)|^2 \dd\omega}{\sum_{c=1}^{C}\sum_{i=1}^{K_{\text{r}}} \int_0^\infty i^2 |\hat{r}^{n+1}_{i,c}(\omega)|^2 \dd\omega} $
        \State \hspace{1em}$\omega_{\text{h}}^{n+1} \gets \frac{\sum_{c=1}^{C}\sum_{j=1}^{K_{\text{h}}} j \int_0^\infty \omega |\hat{h}^{n+1}_{j,c}(\omega)|^2 \dd\omega}{\sum_{c=1}^{C}\sum_{j=1}^{K_{\text{h}}} \int_0^\infty j^2 |\hat{h}^{n+1}_{j,c}(\omega)|^2 \dd\omega} $
        \State 6) Update Lagrangian multiplier $\lambda_c$
        \For{$c=1, 2, \ldots, C$}        
        \State \hspace{1em}$\hat{\lambda}_c^{n+1}(\omega) \gets \hat{\lambda}_c^n(\omega) + \eta \left( \hat{x}_c(\omega) - \hat{x}_{\mathrm{model}, c}^{n+1}(\omega) \right)$
      \EndFor
    \Until Convergence $\frac{\sum_{c=1}^{C} \left\| \hat{x}_{\mathrm{model}, c}^{n+1}(\omega) - \hat{x}_{\mathrm{model}, c}^{n}(\omega) \right\|}{\sum_{c=1}^{C} \left\| \hat{x}_{\mathrm{model}, c}^{n}(\omega) \right\|} < \epsilon_{\mathrm{tol}}$
  \end{algorithmic}
\end{algorithm}

\subsubsection{Instrinstic Mode Function Update}
The optimization procedure for the respiratory mode is given below, considering the terms in the objective function that depend solely on $\hat{r}_{c,i}(\omega)$:
\begin{align}
  \hat{r}^{n+1}_{c,i}(\omega) = \underset{\hat{r}_{c,i}(\omega)}{\operatorname{argmin}} \left\{ \hat{J}_{\rr}\!+\!\left\| \hat{x}_c(\omega)\!-\!\hat{x}_{\mathrm{model}, c}(\omega)\!+\!\frac{\hat{\lambda}_c(\omega)}{2} \right\|^2\right\},
  \label{eq:opt_imf_r}
\end{align}
where $n$ denotes the iteration index in the optimization process.

The analytical solution is derived by determining the point where the gradient of Eq.~\eqref{eq:opt_imf_r} becomes zero.
\begin{align}
  \hat{r}^{n+1}_{c,i}(\omega) &= \frac{\hat{x}_c(\omega) - \hat{x}^n_{\rr, c, i}(\omega)+ \hat{\lambda}^n_c(\omega)/2}{1 + 2\alpha_\rr (\omega-i\omega_\rr)^2},  \nonumber \\ 
\hat{x}^n_{\rr, c, i}(\omega)&=\sum_{i'\neq i}\hat{r}_{c, i'}^{n}(\omega)+\sum_{j=1}^{K_\hh}\hat{h}_{c,j}^n(\omega)+\hat{b}_c^n(\omega)+\hat{g}_c^n(\omega),
\end{align}
where $\sum_{i'\neq i}$ denotes harmonic summation over $i'\in\{1,\ldots,K_\rr\}\setminus\{i\}$, excluding the $i$-th harmonic.

Similarly, the optimization of the heartbeat, trend, and gap components is given as follows:
\begin{align}
  \hat{h}^{n+1}_{c,j}(\omega) &= \underset{\hat{h}_{c,j}(\omega)}{\operatorname{argmin}} \left\{\hat{J}_{\hh}\!+\!\left\| \hat{x}_c(\omega)\!-\!\hat{x}_{\mathrm{model}, c}(\omega)\!+\!\frac{\hat{\lambda}_c(\omega)}{2} \right\|^2  \right\}, \\
    \hat{b}^{n+1}_{c}(\omega) &= \underset{\hat{b}_{c}(\omega)}{\operatorname{argmin}} \left\{\hat{J}_{\bb}\!+\!\left\| \hat{x}_c(\omega)\!-\!\hat{x}_{\mathrm{model}, c}(\omega)\!+\!\frac{\hat{\lambda}_c(\omega)}{2} \right\|^2  \right\}, \\
      \hat{g}^{n+1}_{c}(\omega) &= \underset{\hat{g}_{c}(\omega)}{\operatorname{argmin}} \left\{\hat{J}_{\mathrm{g}}\!+\!\left\| \hat{x}_c(\omega)\!-\!\hat{x}_{\mathrm{model}, c}(\omega)\!+\!\frac{\hat{\lambda}_c(\omega)}{2} \right\|^2  \right\}.
\end{align}

The analytical solutions where the gradient vanishes are then given as follows:
\begin{align}
  \hat{h}^{n+1}_{c,j}(\omega) &= \frac{\hat{x}_c(\omega) - \hat{x}^n_{\hh, c, j}(\omega) + \hat{\lambda}^n_c(\omega)/2}{1 + 2\alpha_\hh (\omega-j\omega_\hh)^2},\nonumber\\ 
  \hat{x}^n_{\hh, c, j}(\omega)&=\sum_{i=1}^{K_\rr}\hat{r}_{c, i}^{n+1}(\omega)+\sum_{j'\neq j}\hat{h}_{c,j'}^n(\omega)+\hat{b}_c^n(\omega)+\hat{g}_c^n(\omega), 
\end{align}
\begin{align}
  \hat{b}^{n+1}_c(\omega) &= \frac{\hat{x}_c(\omega) -  \hat{x}^n_{\bb, c}(\omega) + \hat{\lambda}^n_c(\omega)/2}{1 + 2\alpha_\bb \omega^2},\nonumber\\
  \hat{x}^n_{\bb, c}(\omega)&=\sum_{i=1}^{K_\rr}\hat{r}_{c, i}^{n+1}(\omega)+ \sum_{j=1}^{K_\hh}\hat{h}_{c, j}^{n+1}(\omega)+\hat{g}_c^n(\omega), \label{eq:MPVMD_update_b}
\end{align}
\begin{align}
  \hat{g}^{n+1}_{c}(\omega) &= \frac{\hat{x}_c(\omega) - \hat{x}^n_{\grm, c}(\omega) + \hat{\lambda}^n_c(\omega)/2}{1 + 2\alpha_\grm (\omega-\omega_\grm)^2 + \sum_{i=1}^{K_\rr} 1/\{\alpha_\rr^2(\omega-i\omega_\rr)^4\}}, \nonumber\\
  \hat{x}^n_{\grm, c}(\omega)&=\sum_{i=1}^{K_\rr}\hat{r}_{c, i}^{n+1}(\omega)+ \sum_{j=1}^{K_\hh}\hat{h}_{c, j}^{n+1}(\omega)+\hat{b}_c^{n+1}(\omega). \label{eq:MPVMD_update_g}
\end{align}

\subsubsection{Center Frequency Update}
The respiratory center frequency is updated as follows:
\begin{align}
  \omega_\rr^{n+1} = \underset{\omega_\rr}{\operatorname{argmin}}\quad\left\{\hat{J}_{\rr} + \hat{J}_{\mathrm{g}} \right\}. \label{eq:opt_omega_r}
\end{align}

The stationary condition of Eq.~\eqref{eq:opt_omega_r} is given by
\begin{small}
\begin{align}
  \omega_\rr\!=\!\frac{\sum_{c=1}^{C}\sum_{i=1}^{K_{\text{r}}} i \int_0^\infty \left[\alpha_\rr \omega |\hat{r}_{c,i}(\omega)|^2 - \frac{1}{\alpha_\rr^2(\omega - i\omega_{\text{r}})^5} |\hat{g}_c(\omega)|^2 \right]\!\dd\omega}{\alpha_\rr\sum_{c=1}^{C}\sum_{i=1}^{K_{\text{r}}} \int_0^\infty i^2 |\hat{r}_{c,i}(\omega)|^2 \dd\omega}.
\end{align}
\end{small}
It is difficult to derive a strict analytical solution for $\omega_\rr$ because of the reciprocal of the fifth-order polynomial of $\omega_\rr$ in the second term of the gap component.
As with previous studies~\cite{10.1109/JBHI.2017.2734074, 10.1016/j.sigpro.2020.107610, 10.1109/JBHI.2022.3171554} that incorporated filters into models, when $\alpha_\rr^2$ is sufficiently large, the contribution of the second term in the numerator is considered to be negligible, and a first-term approximation suffices:
\begin{align}
  \omega_\rr^{n+1} & \simeq \frac{\sum_{c=1}^{C}\sum_{i=1}^{K_{\text{r}}} i \int_0^\infty \omega |\hat{r}^{n+1}_{c,i}(\omega)|^2 \dd\omega}{\sum_{c=1}^{C}\sum_{i=1}^{K_{\text{r}}}i^2\int_0^\infty|\hat{r}^{n+1}_{c,i}(\omega)|^2 \dd\omega}. \label{eq:propUpadateCFBR}
\end{align}

The heartbeat center frequency is updated as follows:
\begin{align}
  \omega_\hh^{n+1} &= \underset{\omega_\hh}{\operatorname{argmin}}\quad\hat{J}_{\hh}, \\
  \omega_\hh^{n+1} &= \frac{\sum_{c=1}^{C}\sum_{j=1}^{K_{\text{h}}} j \int_0^\infty \omega |\hat{h}^{n+1}_{c,j}(\omega)|^2 \dd\omega}{\sum_{c=1}^{C}\sum_{j=1}^{K_{\text{h}}} j^2 \int_0^\infty |\hat{h}^{n+1}_{c,j}(\omega)|^2 \dd\omega}. \label{eq:propUpadateCFHR}
\end{align}

\subsubsection{Lagrange Multiplier Update}
The Lagrange multiplier is updated as follows:
\begin{align}
  \hat{\lambda}_c^{n+1}(\omega) &= \hat{\lambda}_c^n(\omega) + \eta \left( \hat{x}_c(\omega) - \hat{x}_{\mathrm{model}, c}^{n+1}(\omega) \right),
\end{align}
where $\hat{x}_{\mathrm{model}, c}^{n+1}(\omega)$ is the model signal defined in Eq.~\eqref{eq:modelPropFreq}, which is reconstructed at the $(n+1)$-st iteration.

\begin{figure*}[t]
  \subfloat[\label{fig:VMDRaw}]{
    \includegraphics[width = 1\textwidth,pagebox=cropbox,clip]{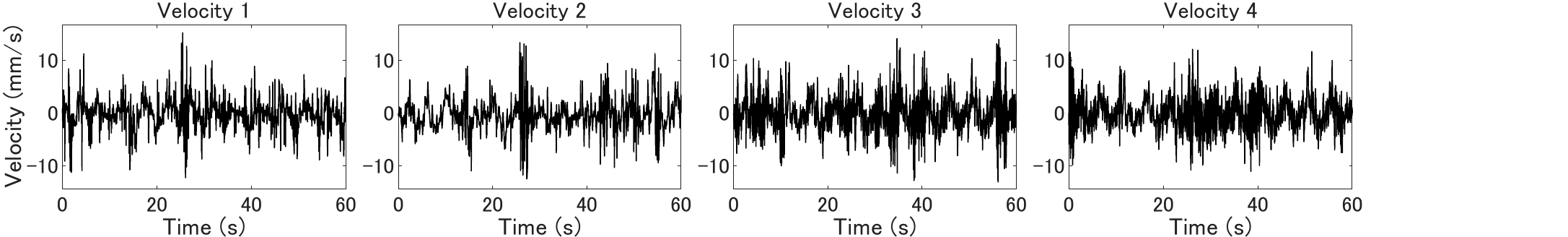}
  }
  \\
  \subfloat[\label{fig:VMDRawFreq}]{
    \includegraphics[width = 1\textwidth,pagebox=cropbox,clip]{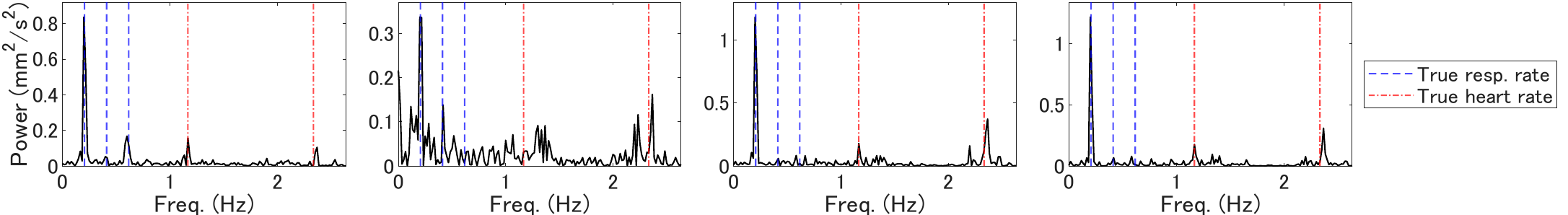}
  }
    \\
  \subfloat[\label{fig:VMDProp1Freq}]{
    \includegraphics[width = 1\textwidth,pagebox=cropbox,clip]{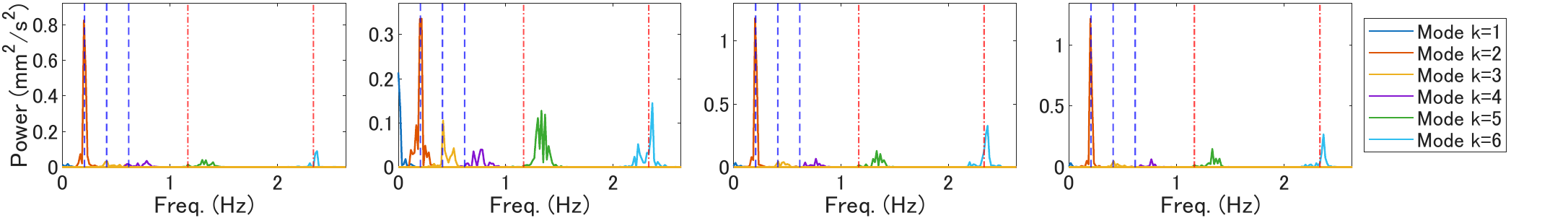}
  }
    \\
  \subfloat[\label{fig:VMDProp3Freq}]{
    \includegraphics[width = 1\textwidth,pagebox=cropbox,clip]{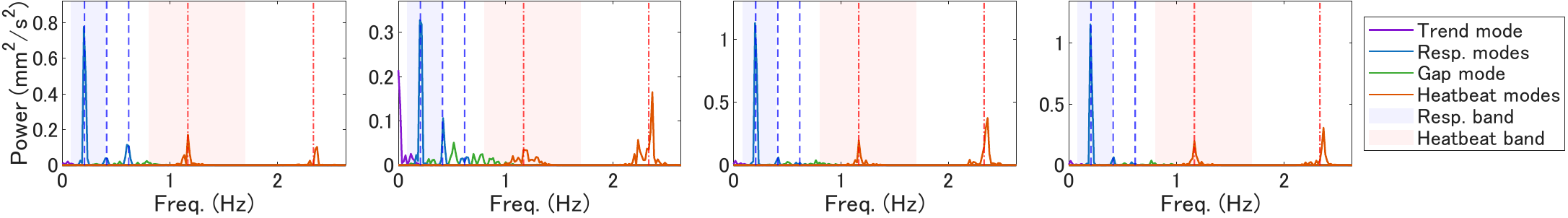}
  }
  \caption{(a) Velocity signals of a human body surface acquired from two radar systems. (b) Corresponding power spectra. (c) Frequency-domain representations of the IMFs using MVMD.  (d) Frequency-domain representation of IMFs using the proposed method. The blue and red dashed lines represent the reference respiration and heartbeat frequencies, respectively, measured using contact sensors.}
  \label{fig:VMDFreqResults}
\end{figure*}

Figure~\ref{fig:VMDFreqResults} shows an example of the mode decomposition results (Subject ID 11 in Section~\ref{sec:Exp2MultiSub}) using MVMD and the proposed method. 
Figure~\ref{fig:VMDRaw} shows the four-channel velocity signals of the human body surface acquired from two radar systems. Figs.~\ref{fig:VMDRawFreq}--\ref{fig:VMDProp3Freq} show the power spectra of the original velocity signals, the modes obtained from MVMD, and the modes obtained from the proposed method, respectively.
Details of the experimental conditions, signal processing procedures, and hyperparameter settings used in each method are provided in Sections~\ref{sec:C4Exp1setup} and \ref{sec:C4MethodParams}.
The blue and red dashed lines in Figs.~\ref{fig:VMDRawFreq}--\ref{fig:VMDProp3Freq} indicate the reference frequencies of respiration and heartbeat using contact sensors. 

As shown in Fig.~\ref{fig:VMDProp1Freq}, MVMD successfully extracts the respiratory mode (the mode index $k=2$, orange) and the second harmonic of the heartbeat signal (mode index $k=6$, cyan), which demonstrates its capability to extract modes commonly shared across all four velocity signals. By contrast, the fundamental heartbeat mode (mode index $k=5$, green) deviates from the reference frequency because of an unwanted component contained in the second-channel velocity.

Figure~\ref{fig:VMDProp3Freq} shows the trend, respiration, gap, and heartbeat modes obtained by using the proposed method, represented in purple, blue, green, and orange, respectively. The number of harmonics was set to $K_\rr=3$ and $K_\hh=3$ in Eq.~\eqref{eq:propModelTimeDomain}, allowing the center frequencies to be estimated by jointly leveraging the fundamental component and harmonics up to the third order. Consequently, the fundamental heartbeat mode (red line) matches the reference value.
Furthermore, the gap component (green) of the second-channel velocity accounts for the unwanted spectral component located between the respiratory and heartbeat frequency bands indicated by the blue and red backgrounds. Because the proposed model imposes harmonic constraints, the reduction in degrees of freedom potentially makes it vulnerable to undesired components. To address this issue, the proposed method incorporates a gap component into the model. This enables the respiratory and heartbeat modes to converge to their respective physiological frequency bands while absorbing gap-band interference, thereby improving the accuracy of frequency estimation.

\subsection{Search for Candidate Initial Values}\label{sec:C3initHR}
As with conventional VMD \cite{10.1109/TSP.2013.2288675}, the proposed algorithm lacks a guarantee of global optimality, which implies that it may converge to a local optimum depending on the initial values.
In particular, the heartbeat signal exhibits high power in its harmonic components, and the center frequency parameter $\omega_\hh$ is updated considering these harmonics, as shown in Eq.~\eqref{eq:propUpadateCFHR}. 
As a result, an integer-multiple ambiguity may arise between the true heartbeat frequency $\omega_\hh^{\mathrm{ref}}$ and the estimated frequency $\hat{\omega}_\hh$.
To address this issue, we prepare $L$ candidate initial values for the heartbeat center frequency, denoted by $\omega_{\mathrm{iniHR}}/2\pi \in \mathcal{S}_{\mathrm{iniHR}} = \{ f_{\mathrm{iniHR}}^{(1)}, \ldots, f_{\mathrm{iniHR}}^{(L)}\}$, and execute the proposed method $L$ times. The result corresponding to the initialization that yields the smallest model error is then selected as follows:
\begin{align}
   l^{*} &= \underset{l}{\operatorname{argmin}}\, \sum_{c=1}^{C} \left\| \hat{x}_{c}(\omega) - \hat{x}^{(l)}_{\mathrm{model}, c}(\omega) \right\|^2,
\end{align}
where $l=1,\hdots,L$ is an index for the initial values, $l^{*}$ denotes the optimal index, and $\hat{x}^{(l)}_{\mathrm{model}, c}(\omega)$ is the model signal obtained from the proposed method using the $l$-th initial value. Note that this initial value search is not performed for the respiration estimation stage, since the fundamental frequency of respiration is relatively easy to extract.

%%%
\section{Experimental Accuracy Evaluation}\label{sec:EXPRESULTS}
We conducted single-person experiments with six participants, described in Section~\ref{sec:C4Exp1setup}, and multiperson experiments with 16 participants, described in Section~\ref{sec:Exp2MultiSub}, to evaluate the effectiveness of the proposed method.
\subsection{Experiment Setup}\label{sec:C4Exp1setup}
\begin{figure}[tb]
  \centering
  \includegraphics[width = 0.95\linewidth,pagebox=cropbox,clip]{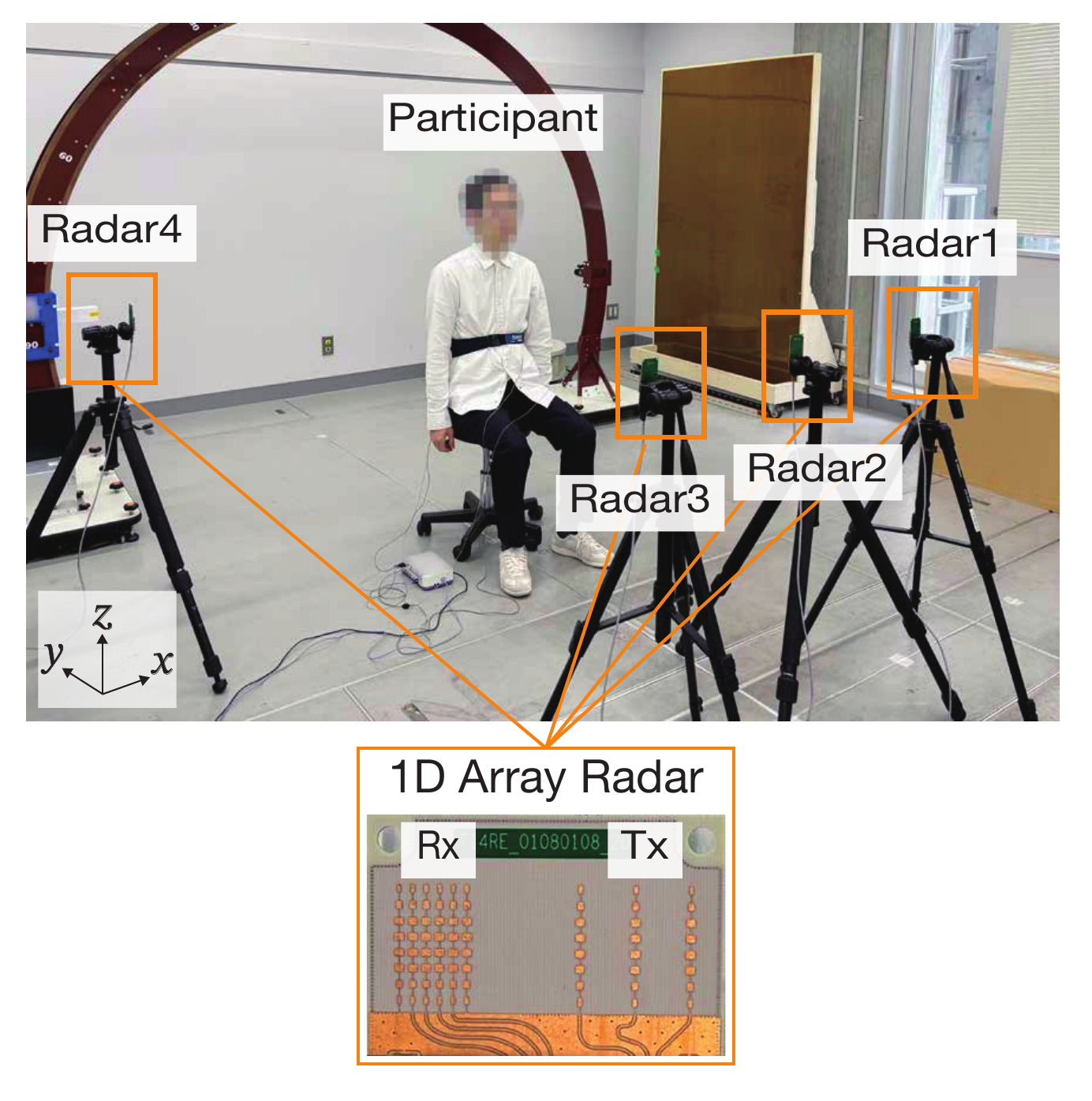}
  \caption{Experimental setup for the single-person scenario using four radar systems.}
  \label{fig:picture_expsetup}
\end{figure}

\begin{figure}[tb]
  \centering
  \subfloat[\label{fig:C4_Conv_radarImage}]{
    \includegraphics[width = 0.48\linewidth,pagebox=cropbox,clip]{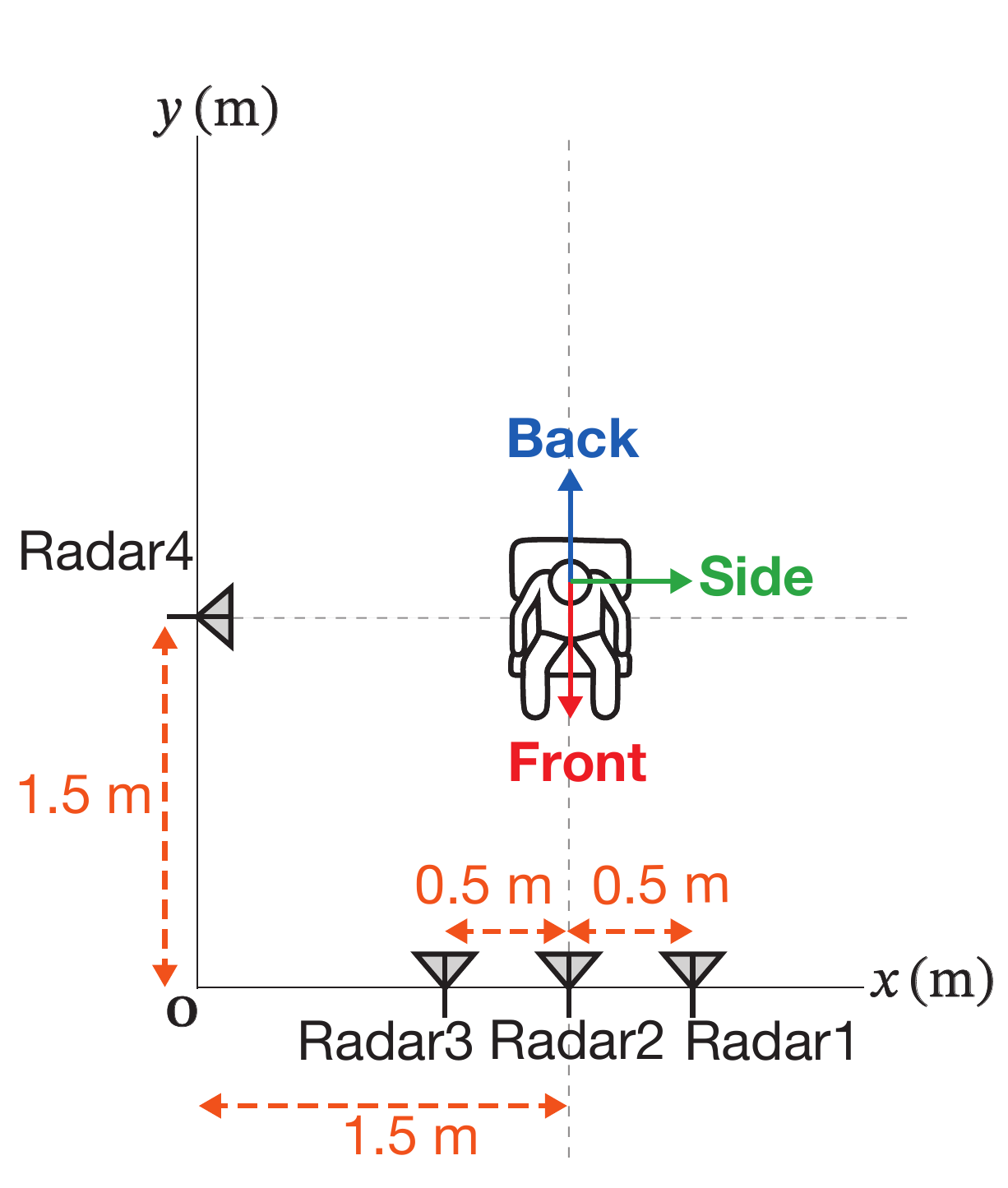}
  }
  \subfloat[\label{fig:C4_Conv_radarImageProposed}]{
    \includegraphics[width = 0.48\linewidth,pagebox=cropbox,clip]{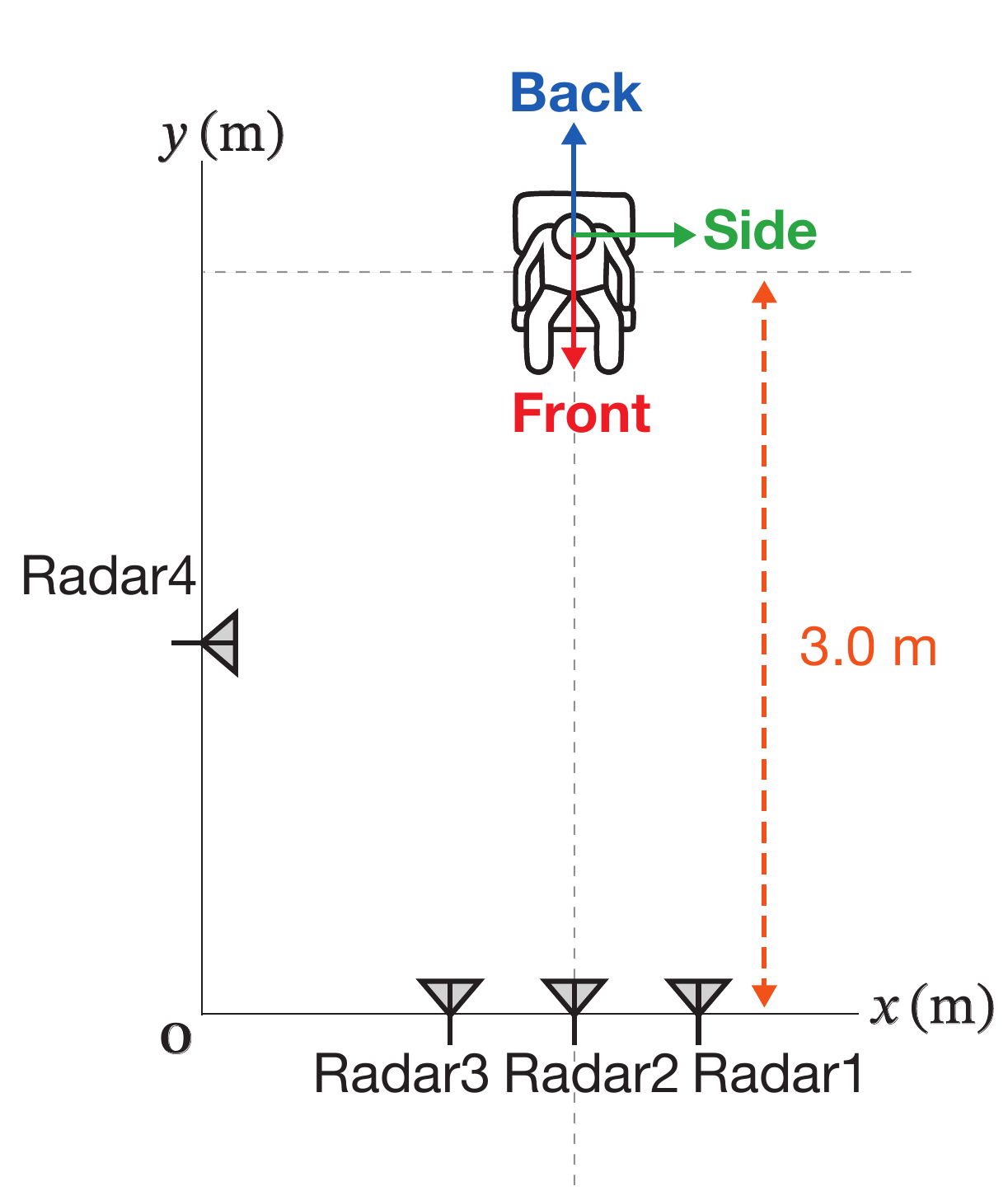}
  }
  \caption{Diagram of the participant and radar arrangement. The participant was positioned at a distance of (a) 1.5~m and (b) 3.0~m.}
  \label{fig:ponchi_expsetup}
\end{figure}

We conducted simultaneous measurements for a single-person scenario using four radar systems.
We used a 1-D array radar device (T14RE\_01080108\_2D; S-Takaya Electronics Industry, Okayama, Japan) with specifications listed in Table~\ref{tbl:radar_system}. 
Figure~\ref{fig:picture_expsetup} shows the experimental environment, and Fig.~\ref{fig:ponchi_expsetup} illustrates the arrangement of the radar systems and participant positions. 
Six healthy male participants with a mean age of 22.8 years (SD $\pm 0.7$) participated in the experiment.
Each participant was seated at two distances (1.5 and 3.0~m) from radar 2, and for each distance, three orientations were tested: facing the front, side, and back relative to radar 2 (Table~\ref{tbl:expCondition}), resulting in a total of 36 datasets (6 participants $\times$ 6 experimental conditions) for the single-person scenario.
The radar height was fixed at 1.0~m. Data were recorded for 120~s under each condition while the participants were instructed to breathe naturally and minimize body movements.
Reference respiratory and cardiovascular data were obtained simultaneously using a respiration belt and an electrocardiogram (ECG) sensor (Polymate Pro MP6000; Miyuki Giken, Tokyo, Japan). 
\begin{table}[tb]
  \centering
  \caption{SPECIFICATIONS OF THE RADAR SYSTEM}
  \begin{threeparttable}[h]
    \begin{tabular}{cc}
      \toprule
      Radar Type                   & 1D                                    \\
      \midrule
      Center frequency            & 79 GHz                                \\
      Bandwidth                    & 3.354 GHz                             \\
      Number of Tx elements                  & 3                                     \\
      Tx element spacing           & 7.6 mm                                \\
      Output power (EIRP)\tnote{*} & 24 dBm                                \\
      Tx element beamwidth         & \multirow{2}{*}{$\pm$35 / $\pm$4 deg.} \\
      (azimuth/elevation)          &                                       \\
      Number of Rx elements                  & 4                                     \\
      Rx element spacing           & 1.9 mm                                \\
      Rx element beamwidth         & \multirow{2}{*}{$\pm$45 / $\pm$4 deg.} \\
      (azimuth/elevation)          &                                       \\
      Sampling frequency           & 100 Hz                                \\
      \bottomrule
    \end{tabular}
    \begin{tablenotes}
      \item[*]EIRP: Equivalent isotropically radiated powers
    \end{tablenotes}
  \end{threeparttable}
  \label{tbl:radar_system}
\end{table}
\begin{table}[tb]
  \centering
  \caption{SINGLE-PERSON EXPERIMENTAL CONDITIONS}
  \begin{tabular}{ccc}
    \toprule
    Condition ID & Position               & Seating direction \\
    \midrule
    C1           & \multirow{3}{*}{1.5~m} & Front  \ang{0}    \\
    C2           &                        & Side  \ang{90}    \\
    C3           &                        & Back  \ang{180}   \\
    \cmidrule{1-3}
    C4           & \multirow{3}{*}{3.0~m} & Front  \ang{0}    \\
    C5           &                        & Side  \ang{90}    \\
    C6           &                        & Back  \ang{180}   \\
    \bottomrule
  \end{tabular}
  \label{tbl:expCondition}
\end{table}

Next, the estimation of the target displacement is described. 
Let $v=1,\ldots,N_{\mathrm{radar}}$ denote the index of the distributed radar systems used for estimation.
By applying the range fast Fourier transform (FFT) and beamforming procedures described in Section~\ref{sec:C2Radar}, we obtain the radar image $I_v(r,\theta,t)$ for the $v$-th radar.
Since each radar image contains multiple echoes from the human body,
we define $N_{\mathrm{tgt}}$ as the number of targets of interest, with their respective coordinates denoted by $(r_{v,c'},\theta_{v,c'})$, $c' = 1, \dots, N_{\mathrm{tgt}}$. 
The line-of-sight velocity of the target is then estimated by demodulating the corresponding signal using the differentiate-and-cross-multiply (DACM) algorithm \cite{10.1109/SIBIRCON.2008.4602570, 10.1109/TIM.2013.2277530}:
\begin{align}
  \dot{d}_{v, c'}(t) &= \frac{\lambda_{\mathrm{radar}}}{4\pi}\frac{ s^{\Icomp}_{v, c'}(t)\dot{s}^{\Qcomp}_{v, c'}(t)-\dot{s}^{\Icomp}_{v, c'}(t)s^{\Qcomp}_{v, c'}(t) }{ \left( s^{\Icomp}_{v, c'}(t) \right)^2 + \left( s^{\Qcomp}_{v, c'}(t) \right)^2 }, \\
  s^{\Icomp}_{v, c'}(t) &= \mathrm{Re}\left[\tilde{I}_{v}(r_{v,c'}, \theta_{v,c'}, t)\right], \\
  s^{\Qcomp}_{v, c'}(t) &= \mathrm{Im}\left[\tilde{I}_{v}(r_{v,c'}, \theta_{v,c'}, t)\right],
\end{align}
where $\lambda_{\mathrm{radar}}$ is the wavelength, $s^{\Icomp}_{v, c'}(t)$ and $s^{\Qcomp}_{v, c'}(t)$ are the in-phase (I) and quadrature (Q) components after signal separation, respectively, and the overdot denotes the time derivative. We deine $\tilde{I}_{v}(r,\theta,t) = I_v(r,\theta,t) - 1/T\int_0^{T} I_v(r,\theta,t) \, \mathrm{d} t$ to be the complex signal after subtracting the static component over the measurement duration $T$. 
A Hampel filter (window length of 2~s and an outlier threshold of $4\sigma$) is then applied to the velocity signals.
As a result, the velocity signal vector consisting of $C=N_{\mathrm{radar}}N_{\mathrm{tgt}}$ channels is obtained as
\begin{equation}
\dot{\bm d}(t)=[\dot d_{1,1}(t),\ldots,\dot d_{N_{\mathrm{radar}},N_{\mathrm{tgt}}}(t)]^\top.
\end{equation}
In this study, $N_{\mathrm{radar}}$ was determined by a specific combination of radar IDs 1--4, the details of which are presented in Section~\ref{sec:C4MethodParams}. 
The parameter $N_{\mathrm{tgt}}$ was set to two for each radar since a limited number of body parts are observable from a single viewpoint.
The target coordinates were manually selected based on the echo power in the radar images. To ensure a fair comparison, identical coordinates were applied across all evaluated methods.

It should be noted that previous studies have proposed advanced range-bin selection methods (i.e., coordinate selection in this paper), which rely on metrics other than echo power, such as phase fluctuations \cite{10.1109/JIOT.2021.3075167, 10.1109/RadarConf2351548.2023.10149752}, autocorrelation \cite{10.1109/ACCESS.2021.3132608, 10.1109/JIOT.2021.3128548, 10.1109/JIOT.2024.3449408}, or the spectral power of physiological signals \cite{10.1145/3589347}.
Although these various adaptive selection methods are promising for improving estimation accuracy, this study mainly focuses on the development of the multivariate signal processing method, which integrates the different human body displacements acquired from distributed radar systems.
These velocity signals $\dot{\bm{d}}(t)$ were used as the input to both the conventional and proposed methods.

\subsection{Parameters of the Proposed and Conventional Methods}\label{sec:C4MethodParams}
We refer to the method using univariate VMD as ``Conv1,’’ the method using MVMD as ``Conv2,’’ and the proposed MPVMD method as ``Prop.’’ 
In Conv1, VMD is applied individually to each of the two target velocity signals acquired by each of radars 1--4. Therefore, the input to the algorithm is a univariate signal with a single channel ($C=1$). The number of data samples for each radar is $N'=72$ (Subjects: 6 $\times$ Conditions: 6 $\times$ $N_{\mathrm{tgt}}:2$) in Conv1.
In contrast, Conv2 and Prop are evaluated using the radar combinations (1,3), (2,4), and (1,3,4). These combinations correspond to $N_{\mathrm{radar}}=2$, 2, and 3 radar systems, respectively. Consequently, the input velocity signal $\dot{\bm{d}}(t)$ consists of $C=4$, 4, and 6 channels, respectively. For each radar combination, the number of data samples is $N'=36$ (Subjects: 6 $\times$ Conditions: 6) for both Conv2 and Prop.

The hyperparameters of the proposed method $\alpha_\rr, \alpha_\hh, \alpha_\bb, \alpha_\grm, K_\rr, K_\hh, \eta, \text{and } \epsilon_{\mathrm{tol}}$ were set as follows. The narrow-bandwidth parameters for respiration and heartbeat were set to $\alpha_\rr=10\ \text{s}^2$ and $\alpha_\hh=1\ \text{s}^2$, the number of harmonics to $K_\rr=3$ and $K_\hh=3$, the update step size for the Lagrange multiplier to $\eta=10^{-2}$, and the threshold for convergence tolerance to $\epsilon_{\mathrm{tol}}=10^{-8}$. Furthermore, the trend and gap parameters $\alpha_\bb$ and $\alpha_\grm$ were determined by the half-bandwidth of the filter $H_{(\cdot)}(\omega)$ in Eqs.~\eqref{eq:MPVMD_update_b} and \eqref{eq:MPVMD_update_g}.
\begin{equation}
  H_{(\cdot)}(\omega) = \frac{1}{1+ 2\alpha_{(\cdot)} \varDelta\omega_{(\cdot)}^2},
\end{equation}
where $\varDelta\omega_{(\cdot)} = \omega-\omega_{(\cdot)}$ is the deviation from the center frequency. Considering the respiration and heartbeat frequency bands, we set $\alpha_\bb \simeq 0.820\ \text{s}^2$ and $\alpha_\grm \simeq 0.131 \ \text{s}^2$, which are the values for which the power $|H(\omega)|^2$ is reduced by half at $\varDelta\omega_{\bb}/2\pi=0.08$ Hz and $\varDelta\omega_{\grm}/2\pi=0.2$ Hz, respectively.
The initial value of the respiration center frequency $\omega_{\mathrm{iniRR}}$ in Algorithm~\ref{alg:PropMPVMD} was selected as the frequency corresponding to the maximum power across all channels within the respiration frequency band $\mathcal{E}_{\mathrm{RR}}$ of the velocity power spectral density. 
The initial values for the heartbeat, $\omega_{\mathrm{iniHR}}$, were determined using the method described in Section~\ref{sec:C3initHR}, with $\mathcal{S}_{\mathrm{iniHR}} = \{1, 1.25, 1.5\}$ Hz, where the number of candidate initial values was $L=3$.

The hyperparameters for univariate VMD (Conv1) were set to $K=10, \alpha=1 \ \text{s}^2, \eta=10^{-2}$, and $\epsilon_{\mathrm{tol}}=10^{-8}$. The initial values for the center frequencies were determined based on the power spectral density of the velocity signal, and an initial value search similar to the one described in Section~\ref{sec:C3initHR} was conducted for the proposed method.
In normal VMD, it is necessary to select the appropriate respiration and heartbeat modes from the $K$ extracted modes. In this study, these modes were determined based on prior estimation of the center frequencies for respiration and heartbeat.
\begin{align}
  k^*_\rr&= \underset{k}{\textrm{argmin}}\,\left| \omega_k - 2\pi\hat{f}_{\mathrm{preRR}} \right|,\,\hat{f}_{\mathrm{preRR}} = \underset{f\in \mathcal{E}_{\mathrm{RR}}}{\textrm{argmax}} \, D(f),\\
  k^*_\hh&= \underset{k}{\textrm{argmin}}\,\left| \omega_k - 2\pi\hat{f}_{\mathrm{preHR}} \right|, \, \hat{f}_{\mathrm{preHR}} = \underset{f\in \mathcal{E}_{\mathrm{HR}}}{\textrm{argmax}} \; D(f),
\end{align}
where $k^*_\rr$ and $k^*_\hh$ are indices of the modes for respiration and heartbeat, respectively, $D(f)$ represents the power spectral density of the velocity signal $\dot{d}(t)$, and $\hat{f}_{\mathrm{preRR}}$ and $\hat{f}_{\mathrm{preHR}}$ are prior estimated frequencies for respiration and heartbeat, respectively. The final estimated center frequencies for respiration and heartbeat were then determined as $(\hat{\omega}_\rr,\,\hat{\omega}_\hh) = (\omega_{k^*_\rr},\, \omega_{k^*_\hh})$.

The hyperparameters $K, \alpha, \eta, \text{and } \epsilon_{\mathrm{tol}}$ for MVMD (Conv2) are respectively identical to those of Conv1 and were set to the same values. For the selection of the respiration and heartbeat modes, the prior estimated frequencies were chosen as the frequencies corresponding to the maximum power within the respective respiration and heartbeat frequency bands of the multivariate signals.

\subsection{Accuracy Evaluation Metrics}
The accuracy of the estimated center frequencies for respiration and heartbeat, $\hat{f}_{p}=\hat{\omega}_p/2\pi$ ($p \in \{\rr, \hh\}$), obtained from each method was evaluated using the mean absolute error (MAE) $\epsilon_p$, the mean absolute percentage error (MAPE) $\tilde{\epsilon}_p$, and the success rate $\alpha_p$.
They are defined as
\begin{align}
  \epsilon_p & = \frac{T_{\mathrm{minute}}}{N'}\sum_{n'=1}^{N'}\left|\hat{f}_{p, n'} - f^{\mathrm{ref}}_{p, n'}\right|,\\
  \tilde{\epsilon}_p  & = \frac{1}{N'}\sum_{n'=1}^{N'}\frac{\left|\hat{f}_{p, n'} - f^{\mathrm{ref}}_{p, n'}\right|}{f^{\mathrm{ref}}_{p, n'}}, \\
  \alpha_p & = \frac{|\mathcal{I}_{p}|}{N'}, \\
  \mathcal{I}_p &= \left\{n' \in \left\{1, \cdots, N'\right\}\,\middle|\,\frac{\left|\hat{f}_{p, n'} - f^{\mathrm{ref}}_{p, n'}\right|}{f^{\mathrm{ref}}_{p, n'}}  < \theta^{\mathrm{thr}}_{p}\right\},
\end{align}
where $n'$ is the data index, $N'$ is the number of data samples, $f^{\mathrm{ref}}_{\rr, n'}$ and $f^{\mathrm{ref}}_{\hh, n'}$ are the reference values obtained from the respiration belt and ECG sensor, respectively, and $T_{\mathrm{minute}}=60$ s is a constant used to convert the unit into beats per minute (bpm). Here, $|\mathcal{I}_{p}|$ represents the number of data points that satisfy the success rate threshold $\theta^{\mathrm{thr}}_{p}$, which was set to $\theta^{\mathrm{thr}}_{p}=10\%$ for both respiration and heartbeat.

\subsection{Exp1: Performance Evaluation in Single-Person Scenario}
\begin{figure}[tb]
  \centering
  \includegraphics[width = 1.0\linewidth,pagebox=cropbox,clip]{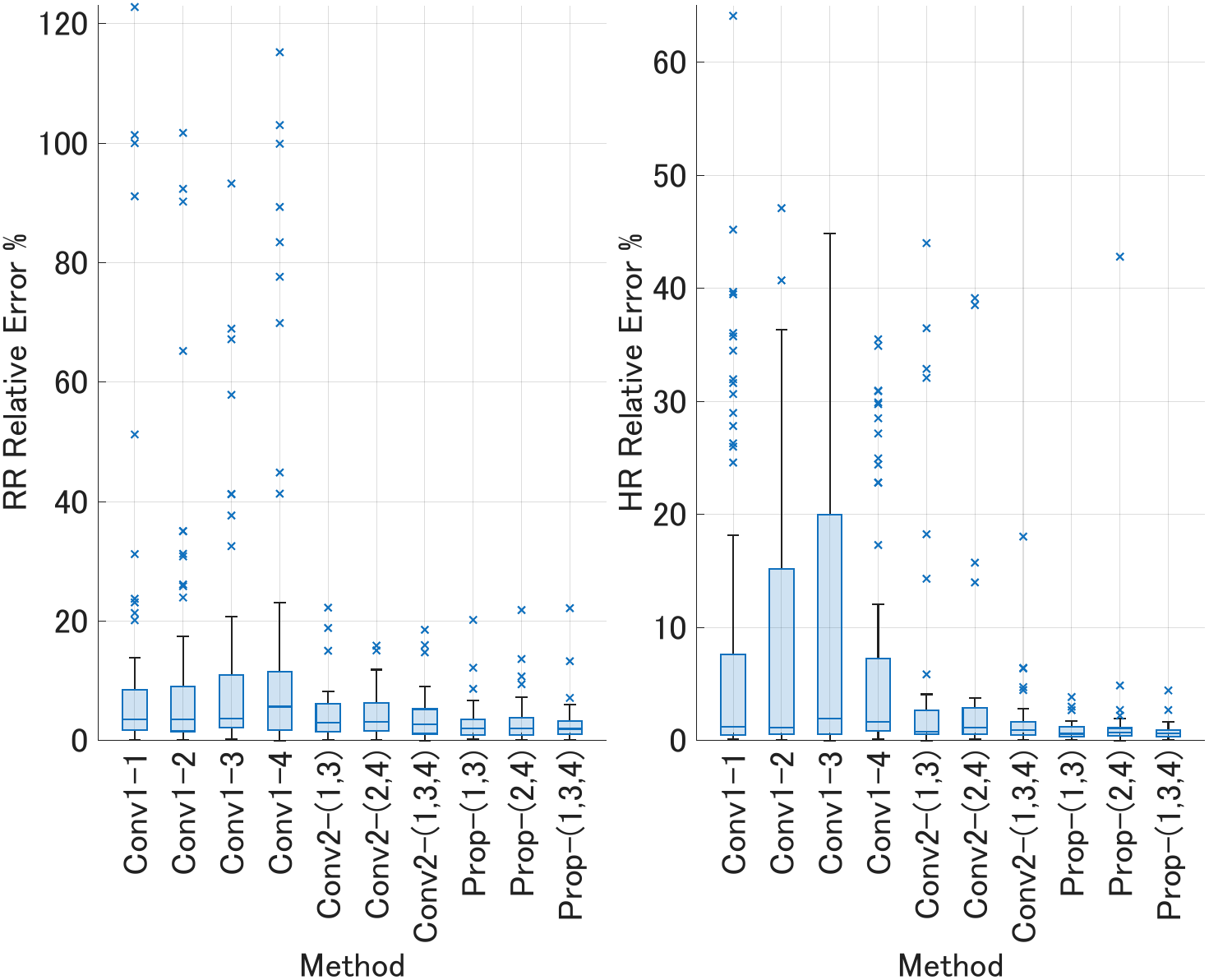}
  \caption{Boxplots of the relative error for respiration and heartbeat estimation across all experiment conditions in the single-person scenario.}
  \label{fig:BoxAllcondExp1}
\end{figure}

\begin{figure*}[t]
  \centering
  \subfloat[\label{fig:BoxEachcondExp1BR}]{
    \includegraphics[width = 1.0\textwidth,pagebox=cropbox,clip]{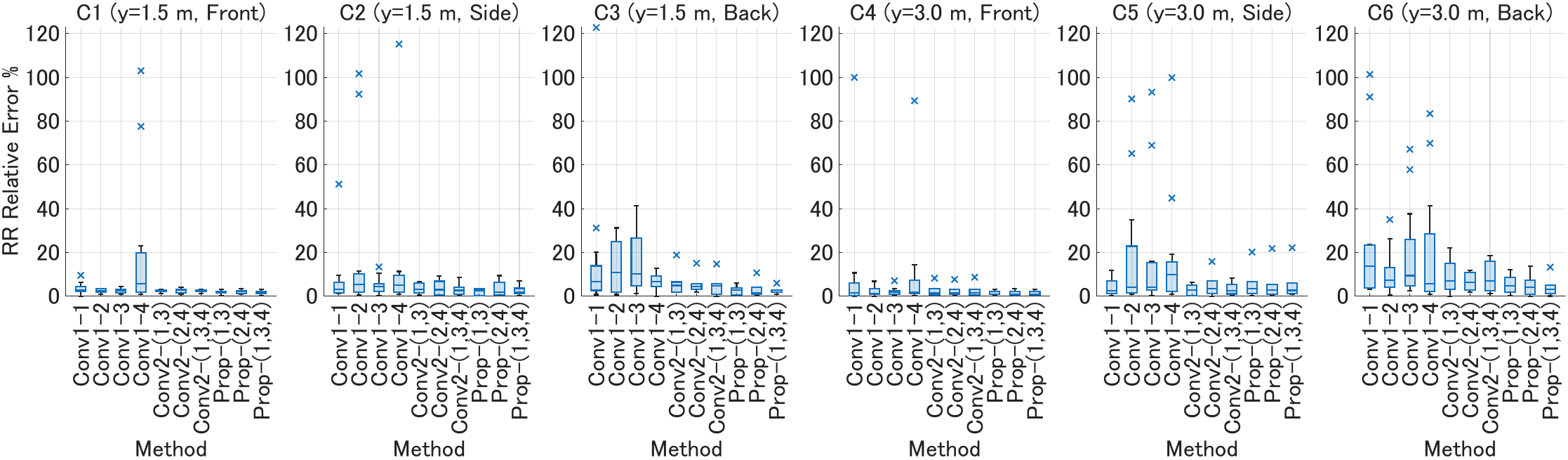}
  }
  \\
  \subfloat[\label{fig:BoxEachcondExp1HR}]{
    \includegraphics[width = 1.0\textwidth,pagebox=cropbox,clip]{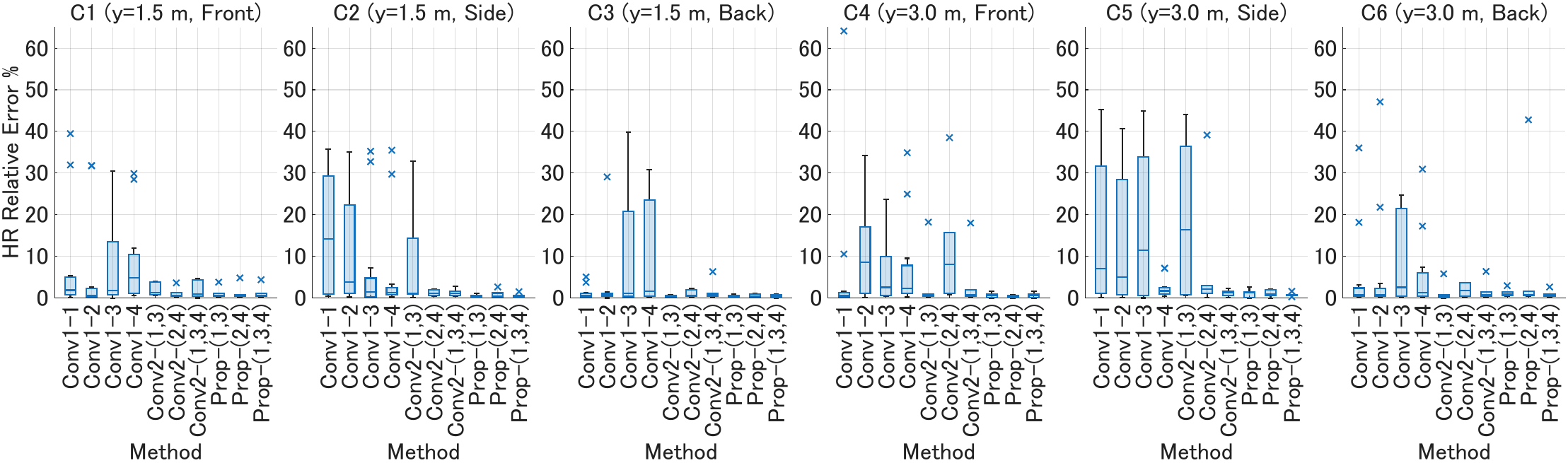}
  }
  \caption{Boxplots of the relative error under each experimental condition in the single-person scenario for (a) respiration and (b) heartbeat estimation.}
  \label{fig:BoxEachcondExp1}
\end{figure*}

Table~\ref{tbl:metricsAllConditionExp1} summarizes the mean performance under all six conditions (two distances $\times$ three orientations). Figure~\ref{fig:BoxAllcondExp1} shows the box plots of the relative error $|\hat{f}_{p, n'} - f^{\mathrm{ref}}_{p, n'}|/f^{\mathrm{ref}}_{p, n'}$ for all experimental conditions combined, while Fig.~\ref{fig:BoxEachcondExp1} presents the box plots for each individual experimental condition.

\begin{table}[tb]
    \centering
      \caption{RESULTS OF ALL PARTICIPANTS IN SINGLE-PERSON SCENARIO}
  \begin{threeparttable}[h]
\begin{tabular}{cccccc}
  \toprule
     &\multirow{2}{*}{Method}  & \multirow{2}{*}{Radar IDs}  & MAE & \multirow{2}{*}{MAPE \%} & \multirow{2}{*}{Success Rate \%}    \\
    &                              &                               &(bpm) & & \\
  \midrule
  \multirow{10}{*}{RR} & \multirow{4}{*}{Conv1} & 1         & 1.35      & 11.58          & 80.6        \\
                       &                        & 2         & 1.28      & 11.29          & 76.4        \\
                       &                        & 3         & 1.28      & 10.62          & 73.6        \\
                       &                        & 4         & 1.71      & 15.19          & 68.1        \\
  \cmidrule{2-6}
                       & \multirow{3}{*}{Conv2} & (1, 3)    & 0.53      & 4.50           & 91.7        \\
                       &                        & (2, 4)    & 0.54      & 4.44           & 88.9        \\
                       &                        & (1, 3, 4) & 0.51      & 4.29           & 91.7        \\
  \cmidrule{2-6}
                       & \multirow{3}{*}{Prop}  & (1, 3)    & 0.39      & 3.23           & 94.4        \\
                       &                        & (2, 4)    & 0.42      & 3.50           & 91.7        \\
                       &                        & (1, 3, 4) & 0.37      & 3.10           & 94.4        \\
  \midrule
  \multirow{10}{*}{HR} & \multirow{4}{*}{Conv1} & 1         & 6.48      & 8.74           & 76.4        \\
                       &                        & 2         & 6.48      & 8.44           & 72.2        \\
                       &                        & 3         & 7.44      & 9.65           & 69.4        \\
                       &                        & 4         & 5.14      & 6.94           & 80.6        \\
  \cmidrule{2-6}
                       & \multirow{3}{*}{Conv2} & (1, 3)    & 4.44      & 5.81           & 83.3        \\
                       &                        & (2, 4)    & 3.21      & 4.25           & 88.9        \\
                       &                        & (1, 3, 4) & 1.37      & 1.93           & 97.2        \\
  \cmidrule{2-6}
                       & \multirow{3}{*}{Prop}  & (1, 3)    & 0.66      & 0.87           & 100       \\
                       &                        & (2, 4)    & 1.62      & 2.05           & 97.2        \\
                       &                        & (1, 3, 4) & 0.62      & 0.81           & 100       \\
  \bottomrule
\end{tabular}
      \begin{tablenotes}
      \item[]RR: respiratory rate, HR: heart rate, MAE: mean absolute error, MAPE: mean absolute percentage error, bpm: beats per minute
    \end{tablenotes}
  \end{threeparttable}
  \label{tbl:metricsAllConditionExp1}
\end{table}

For respiration estimation, Conv1 resulted in an MAE of 1.41 bpm, a MAPE of 12.17\%, and a success rate of 74.7\% across all radars (radar IDs 1--4). For the radar combination (1, 3, 4), Conv2 based on MVMD achieved an MAE of 0.51 bpm, a MAPE of 4.29\%, and a success rate of 91.7\%, corresponding to a reduction of 7.88 percentage points in MAPE and an increase of 17.0 percentage points in success rate compared with Conv1. Using the same radar combination, Prop yielded an MAE of 0.37 bpm, a MAPE of 3.10\%, and a success rate of 94.4\%, demonstrating a reduction of 9.07 percentage points in MAPE and an increase of 19.8 percentage points in success rate over Conv1.
As shown in Fig.~\ref{fig:BoxEachcondExp1BR}, under experimental conditions C5 and C6, where the participant was positioned at a distance of $y = 3.0$ m in the side and back body orientations, both Conv2 and Prop substantially reduced the number of respiration estimates with error exceeding $10$\%, resulting in improved overall accuracy.
By contrast, the difference in performance between Conv2 and Prop was relatively small, which suggests that integrating velocity signals from multiple viewpoints is the main factor that contributes to improved accuracy in estimating respiration.

For heartbeat estimation, Conv1 resulted in an MAE of 6.39 bpm, a MAPE of 8.44\%, and a success rate of 74.7\% across all radar systems. In comparison, Conv2 with the radar combination (1, 3, 4) achieved an MAE of 1.37 bpm, a MAPE of 1.93\%, and a success rate of 97.2\%, which represent improvements over Conv1 of 6.52 percentage points in MAPE and 22.6 percentage points in success rate. Prop achieved the highest performance, with an MAE of 0.62 bpm, a MAPE of 0.81\%, and a success rate of 100\%, corresponding to a reduction of 7.63 percentage points in MAPE and an increase of 25.3 percentage points in success rate over Conv1.
As illustrated in Fig.~\ref{fig:BoxEachcondExp1HR}, both Conv2 and Prop reduced the number of estimates with errors exceeding 10\% across all experimental conditions (C1--C6) compared with Conv1. 
A direct comparison between Conv2 and Prop reveals that Prop further reduced outliers for the radar combination (1, 3) under conditions C2 and C5 and for the radar combination (2, 4) under condition C4. 
Furthermore, for the radar combinations (1, 3) and (2, 4), Conv2 achieved heartbeat success rates of 88.3\% and 88.9\%, respectively, whereas Prop achieved success rates  of 100\% and 97.2\%. These findings suggest that the incorporation of harmonic information allowed Prop to achieve high estimation accuracy even when only a limited number of radar viewpoints were available.

To summarize, the results demonstrate that, for both respiration and heartbeat estimation, integrating displacements acquired from multiple radar viewpoints results in higher accuracy than using a velocity signal obtained from a single radar viewpoint.
Furthermore, the results suggest that the incorporation of harmonic information in the proposed method contributes to more accurate estimation of the heartbeat center frequency.

\subsection{Exp2: Performance Evaluation in Multiperson Scenarios}\label{sec:Exp2MultiSub}
\begin{figure}[tb]
  \centering
  \includegraphics[width = 1.0\linewidth,pagebox=cropbox,clip]{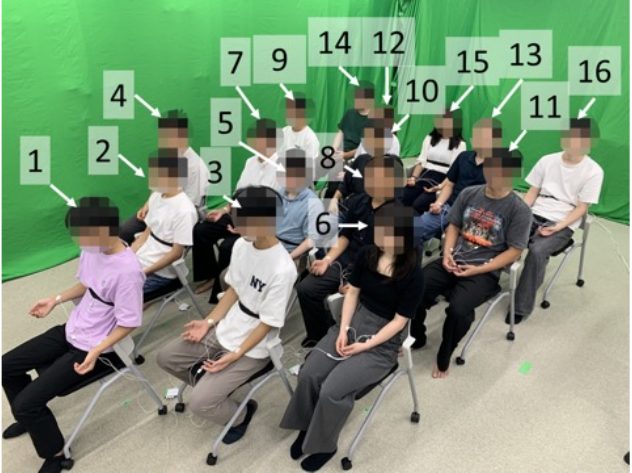}
  \caption{Experimental setup for the multiperson scenario.}
  \label{fig:pictureExpSetupExp2}
\end{figure}

\begin{figure}[tb]
  \centering
  \includegraphics[width = 0.7\linewidth,pagebox=cropbox,clip]{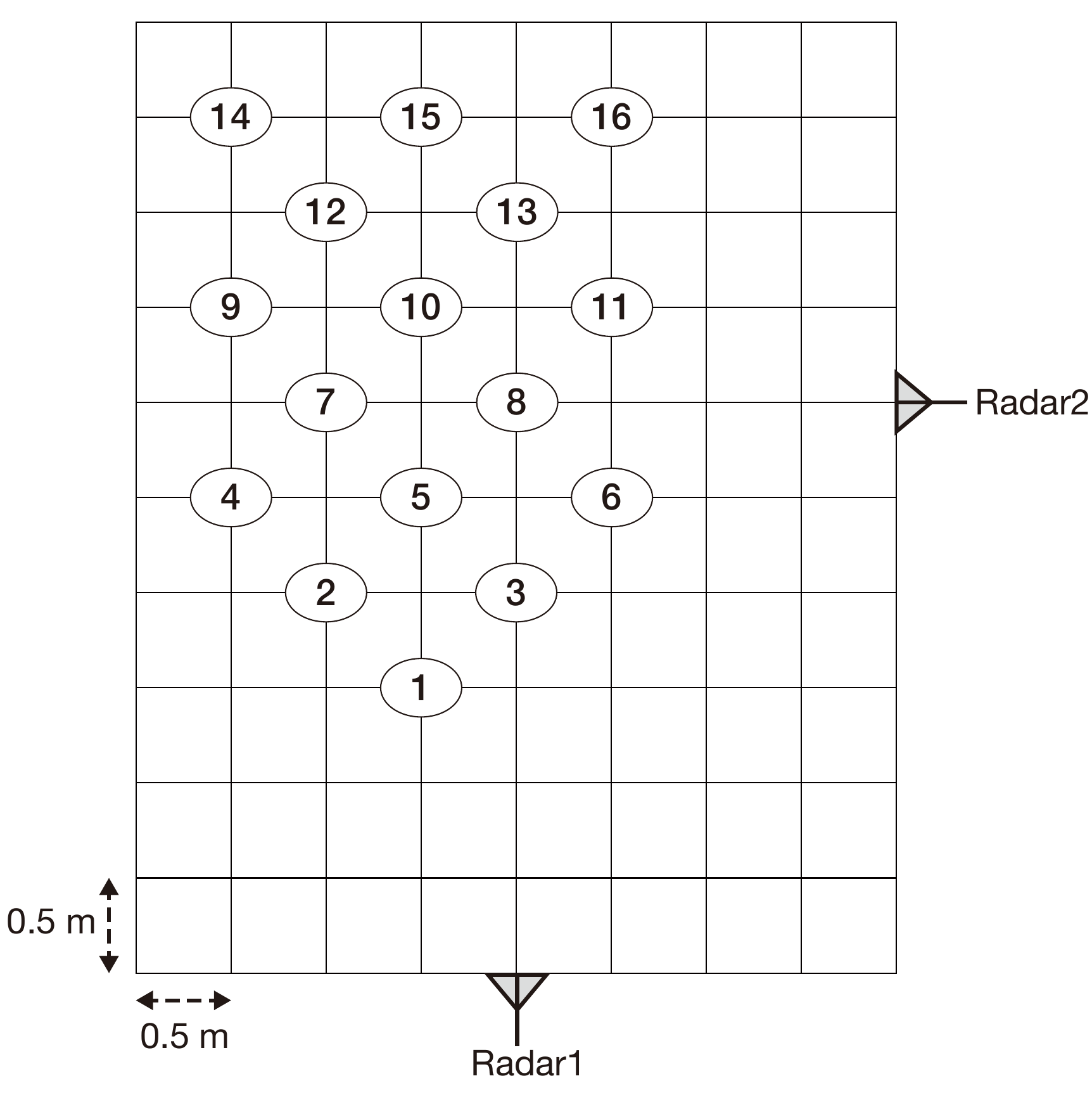}
  \caption{Diagram of subject and radar arrangement in the multi-person scenario.}
  \label{fig:ponchiExpSetupExp2}
\end{figure}

\begin{figure}[tb]
  \centering
  \subfloat[\label{fig:Exp2RadarImage1}]{
    \includegraphics[width = 0.8\linewidth,pagebox=cropbox,clip]{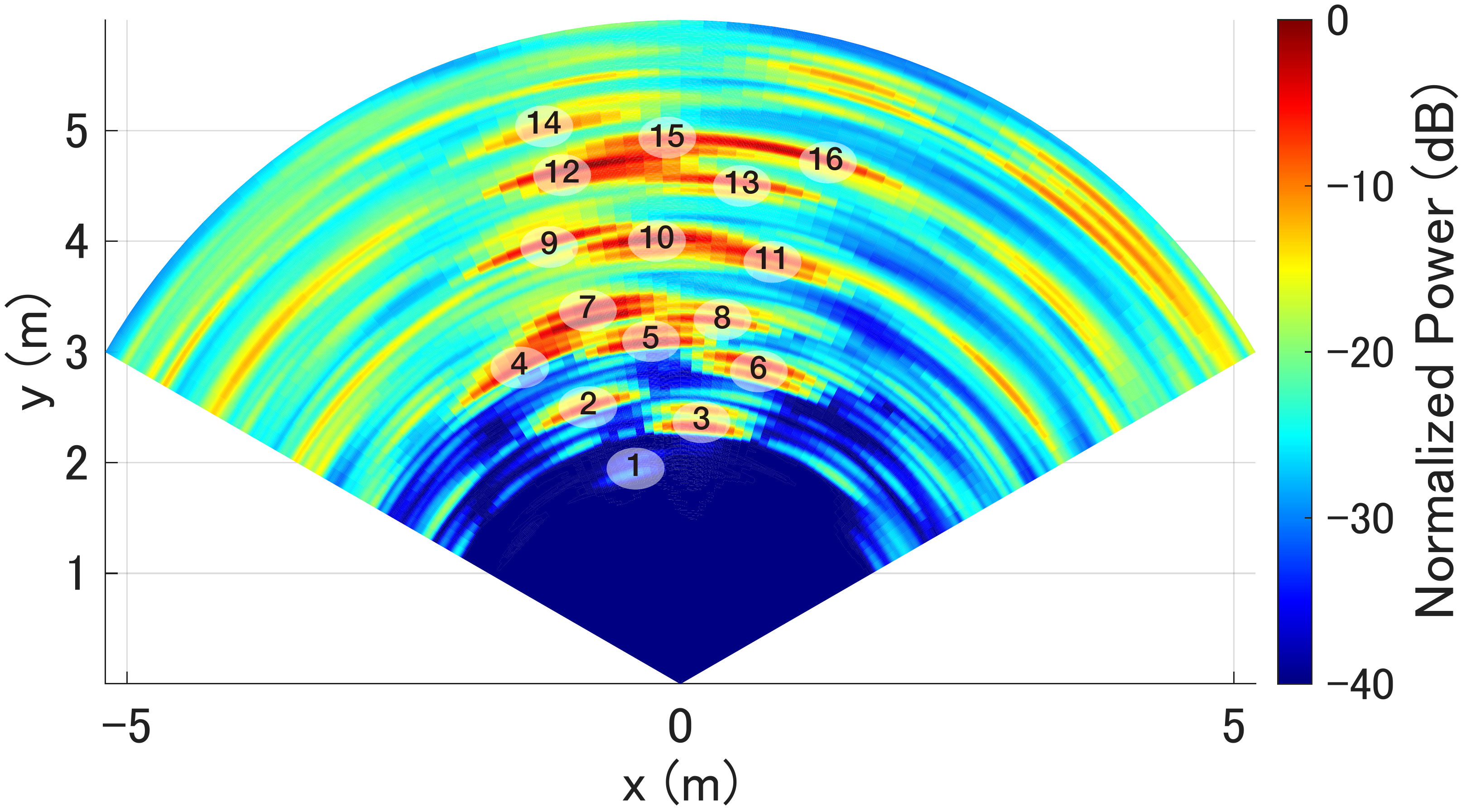}
  }
  \\
  \subfloat[\label{fig:Exp2RadarImage2}]{
    \includegraphics[width = 0.8\linewidth,pagebox=cropbox,clip]{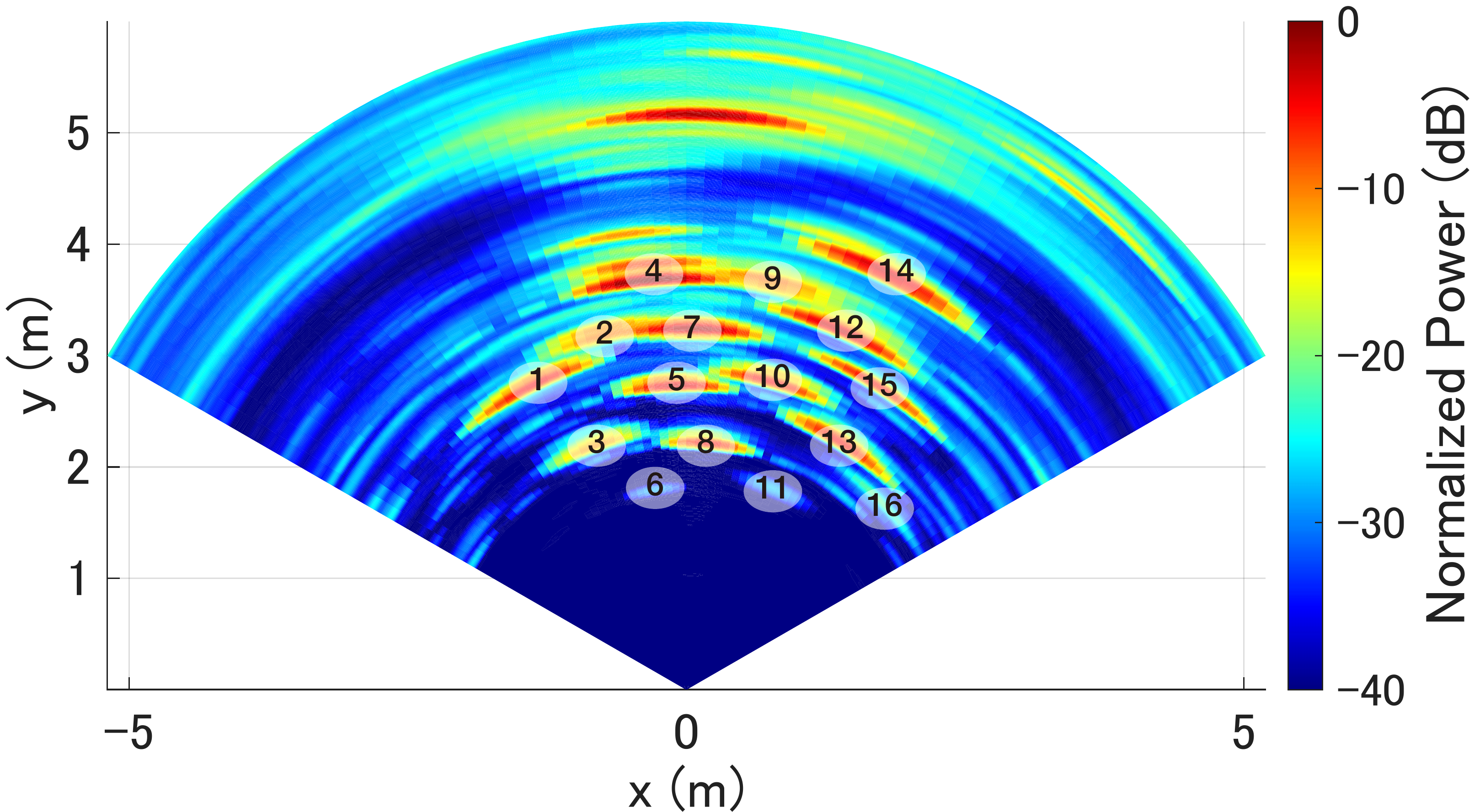}
  }
  \caption{Radar images obtained from (a) radar 1 and (b) radar 2 in the multi-person scenario.}
  \label{fig:radarImageExp2}
\end{figure}

To evaluate the effectiveness of the proposed method in multiperson scenarios, we conducted simultaneous measurement experiments involving 16 participants.
Figure~\ref{fig:pictureExpSetupExp2} illustrates the experimental setup, and Fig.~\ref{fig:ponchiExpSetupExp2} shows the layout of the radar systems and the participants.
The participants consisted of 13 men and 3 women, with a mean age of 22.6 years (SD $\pm 1.6$). All participants were measured simultaneously for 60~s while facing toward radar 1.
The radar systems were configured as in Section~\ref{sec:C4Exp1setup}, with an installation height of 2.0~m and a downward elevation angle of 10 degrees.
The reference signals were acquired using respiration belts and ECG sensors (biosignalsplux; PLUX Wireless Biosignals S.A., Lisbon, Portugal).
The velocity estimation procedures and the hyperparameters for each method were kept exactly the same as in the single-person experiments described in Sections~\ref{sec:C4Exp1setup} and \ref{sec:C4MethodParams}. Figure~\ref{fig:radarImageExp2} shows the radar images acquired from the two radar systems along with the corresponding positions of the participants.

\begin{table}[tb]
  \caption{RESULTS OF ALL PARTICIPANTS IN MULTIPERSON SCENARIO}
  \centering
  \begin{tabular}{cccccc}
  \toprule
     &\multirow{2}{*}{Method}  & \multirow{2}{*}{Radar IDs}  & MAE & \multirow{2}{*}{MAPE \%} & \multirow{2}{*}{Success Rate \%}    \\
    &                              &                               &(bpm) & & \\
                    \midrule
\multirow{4}{*}{RR} & \multirow{2}{*}{Conv1}   & 1        & 1.43      & 16.18          & 81.3        \\
                    &    & 2        & 2.08      & 18.96          & 65.6        \\
                      \cmidrule{2-6}
                    & Conv2  & (1, 2)   & 0.39      & 3.86           & 87.5        \\
                    & Prop  & (1, 2)   & 0.39      & 3.19           & 100       \\
                    \midrule
\multirow{4}{*}{HR} & \multirow{2}{*}{Conv1}    & 1        & 9.30      & 12.59          & 65.6        \\
                    &    & 2        & 10.19     & 12.37          & 62.5        \\
                    \cmidrule{2-6}
                    & Conv2  & (1, 2)   & 8.43      & 10.75          & 56.3        \\
                    & Prop  & (1, 2)   & 3.69      & 6.25           & 87.5        \\
    \bottomrule
  \end{tabular}
  \label{tbl:metricsAllConditionExp2}
\end{table}

\begin{figure}[tb]
  \centering
  \includegraphics[width = 0.95\linewidth,pagebox=cropbox,clip]{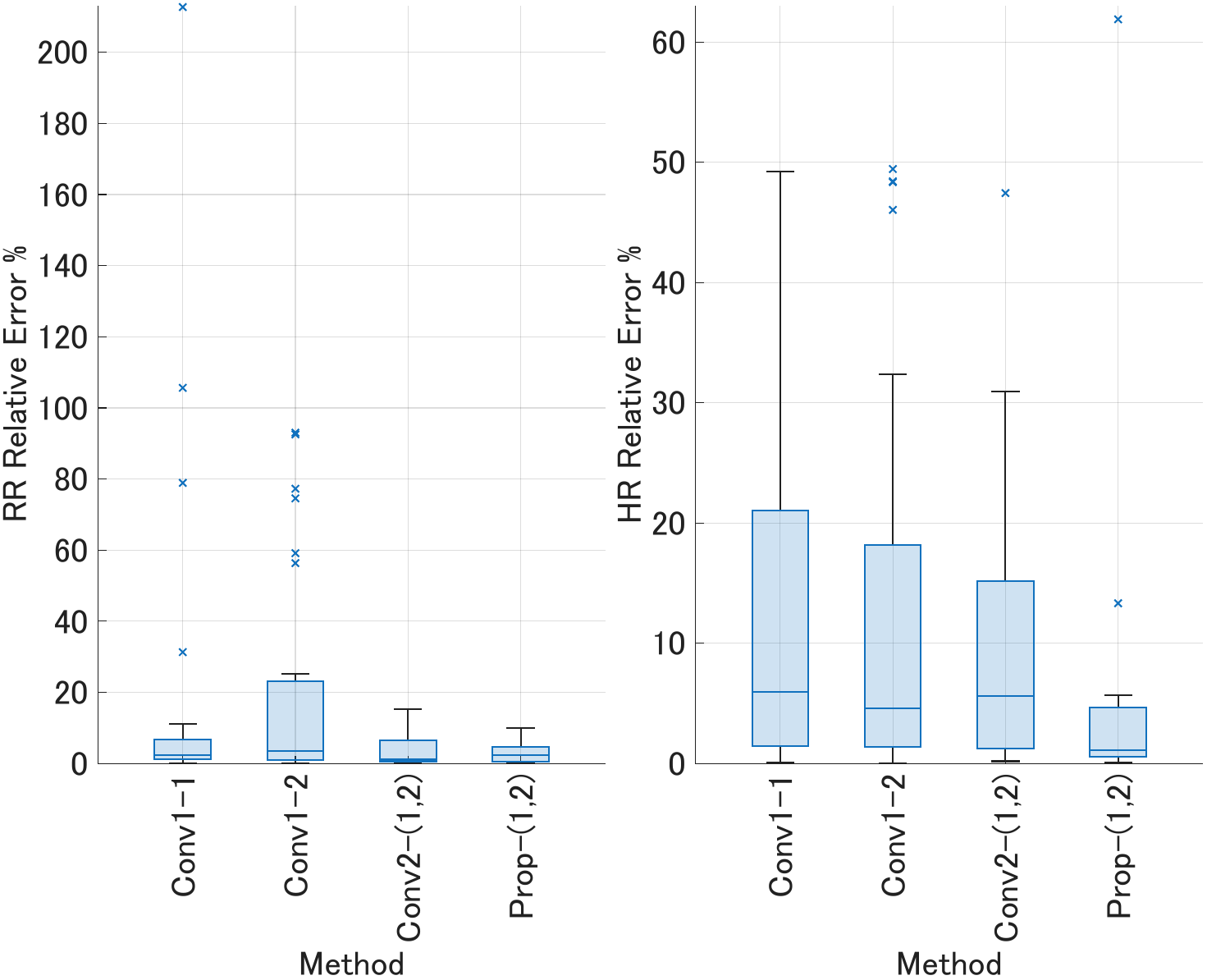}
  \caption{Boxplots of the relative error for respiration and heartbeat estimation in the multi-person scenario.}
  \label{fig:boxAllcondExp2}
\end{figure}

Table~\ref{tbl:metricsAllConditionExp2} summarizes the mean performance across all participants, and Fig.~\ref{fig:boxAllcondExp2} presents the corresponding boxplots.
First, for respiration estimation, the mean performance of Conv1 across both radars (radar IDs 1 and 2) yielded an MAE of 1.75 bpm, a MAPE of 17.57\%, and a success rate of 73.4\%. In comparison, Conv2 resulted in an MAE of 0.39 bpm, a MAPE of 3.86\%, and a success rate of 87.5\%, corresponding to a reduction of 13.71 percentage points in MAPE and an increase of 14.1 percentage points in success rate. Prop resulted in an MAE of 0.39 bpm, a MAPE of 3.19\%, and a success rate of 100\%, corresponding to a reduction of 14.38 percentage points in MAPE and an increase of 26.6 percentage points in success rate compared with Conv1. 
The results demonstrate that Prop achieved better performance than Conv2 in terms of MAPE and success rate, while maintaining a comparable MAE, which confirms the effectiveness of the proposed method.

For heartbeat estimation, Conv1 resulted in an MAE of 9.74 bpm, a MAPE of 12.48\%, and a success rate of 64.1\% across both radars. In contrast, Conv2 yielded an MAE of 8.43 bpm, a MAPE of 10.75\%, and a success rate of 56.3\%, corresponding to a reduction of 1.73 percentage points in MAPE but a decrease of 7.8 percentage points in success rate. Thus, no clear improvement was observed with Conv2, which implies that naive integration of velocity signals acquired from distributed radar systems does not always guarantee sufficient performance.
By contrast, Prop resulted in an MAE of 3.69 bpm, a MAPE of 6.25\%, and a success rate of 87.5\%, corresponding to a reduction of 6.23 percentage points in MAPE and an increase of 23.4 percentage points in success rate compared with Conv1. These results imply that the proposed method can effectively extract both respiratory and heartbeat signals from multiple radar observations even in multi-person scenarios.

\section{Conclusion}
This study proposes a multivariate signal processing approach that extracts physiological signals for robust, non-contact vital sign detection using distributed radar systems under various body-orientation conditions.
The proposed method extends the MVMD framework by incorporating the harmonic structure of physiological signals and the gap component, thereby enabling the extraction of the common center frequency of respiration and heartbeat across the distributed radar systems.
Single-person experiments in which the position and body-orientation were varied were conducted and validated against reference signals obtained via  respiration belt and ECG sensor. The proposed method successfully fused data from three radar systems, yielding success rates of 94.4\% for respiration and 100\% for heartbeat. This represents an improvement over the conventional univariate VMD method, which was outperformed by 19.8 percentage points for respiration and 25.3 percentage points for heartbeat.
Furthermore, we conducted experiments involving 16 participants with two radar systems to evaluate the performance of the proposed method in multi-person scenarios. The proposed method achieved success rates of 100\% for respiration and 87.5\% for heartbeat, improvements of 26.6 and 23.4 percentage points, respectively, over the conventional method.
To summarize, across the experiments in both the single- and multi-person scenarios, the proposed method consistently achieved success rates of over 85\% for both vital signs, which demonstrates its capability for stable respiration and heartbeat signal extraction under diverse body-orientation conditions.
Future work will include the validation in environments with random body movements or various non-sitting postures, excluding sit-down, as well as the development of a real-time implementation of the proposed method.

\section*{Ethics Declarations}
The experimental protocol involving human participants was approved by the Ethics Committee of the Graduate School of Engineering, Kyoto University (permit nos. 202223 and 202408). Informed consent was obtained from all human participants in the study.

\section*{Acknowledgment}
The authors report no conflict of interest.
This work was supported in part by the SECOM Science and Technology Foundation; in part by the Japan Science and Technology Agency under Grant JPMJMS2296 and Grant JPMJSP2110 (Support
for Pioneering Research Initiated by the Next Generation: SPRING); in part by the Japan Society for the Promotion of Science KAKENHI program under Grant 23H01420, Grant 23K26115, and Grant 26K00959; and in part by the New Energy and Industrial Technology Development Organization.
The authors thank Hiromasa Teramachi of Kyoto University for his support in performing radar measurements and analyzing the reference data. We thank Edanz (https://jp.edanz.com/ac) for editing a draft of this manuscript.

\begin{IEEEbiography}[{\includegraphics[width=1in,height=1.25in,clip,keepaspectratio]{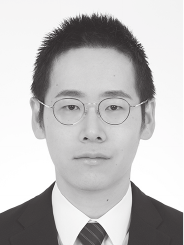}}]
  {Kimitaka Sumi} (Graduate Student Member, IEEE) received B.E. and M.E. degrees from Nagoya University, Nagoya, Japan, in 2020 and 2022, respectively. He is currently pursuing a Ph.D. degree at the Graduate School of Engineering, Kyoto University. His research interests include signal processing and remote sensing.
\end{IEEEbiography}
\begin{IEEEbiography}[{\includegraphics[width=1in,height=1.25in,clip,keepaspectratio]{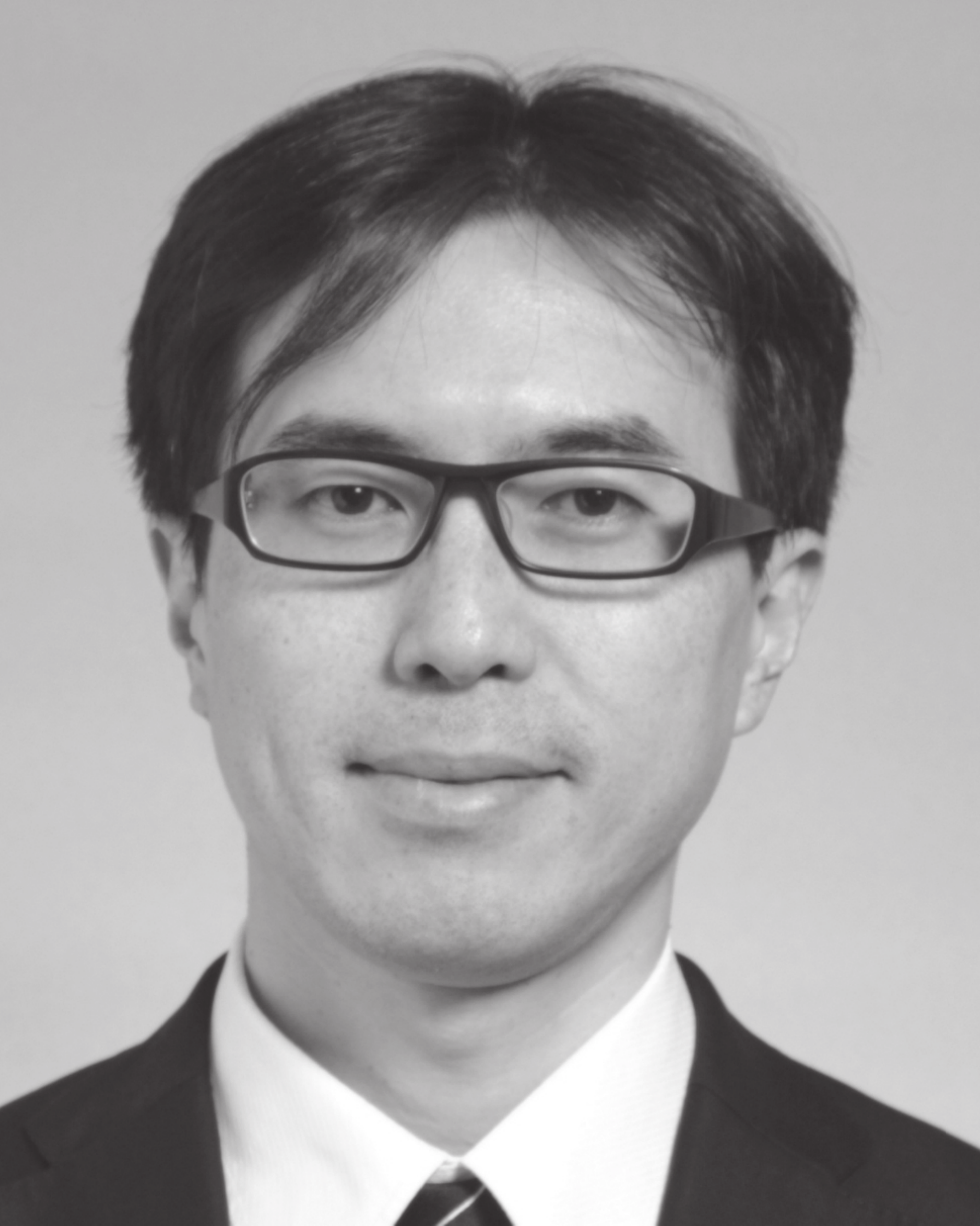}}]
  {Takuya Sakamoto} (Senior Member, IEEE)
  received a B.E. degree in electrical and electronic engineering from Kyoto University, Kyoto, Japan, in 2000 and M.I. and Ph.D. degrees in communications and computer engineering from the Graduate School of Informatics, Kyoto University in 2002 and 2005, respectively. From 2006 through 2015, he was an Assistant Professor at the Graduate School of Informatics, Kyoto University. From 2011 through 2013, he was also a Visiting Researcher at Delft University of Technology, Delft, the Netherlands. From 2015 until 2019, he was an Associate Professor at the Graduate School of Engineering, University of Hyogo, Himeji, Japan. In 2017, he was also a Visiting Scholar at the University of Hawaii at Manoa, Honolulu, HI, USA. From 2019 until 2022, he was an Associate Professor at the Graduate School of Engineering, Kyoto University. From 2018 through 2022, he was a PRESTO researcher of the Japan Science and Technology Agency, Japan. Since 2022, he has been a Professor at the Graduate School of Engineering, Kyoto University. His current research interests lie in wireless human sensing, radar signal processing, and radar measurement of physiological signals.

  Prof. Sakamoto was a recipient of the Best Paper Award from the International Symposium on Antennas and Propagation (ISAP) in 2004, the Young Researcher's Award from the Institute of Electronics, Information and Communication Engineers of Japan (IEICE) in 2007, the Best Presentation Award from the Institute of Electrical Engineers of Japan in 2007, the Best Paper Award from the ISAP in 2012, the Achievement Award from the IEICE Communications Society in 2015, 2018, and 2023, the Achievement Award from the IEICE Electronics Society in 2019, the Masao Horiba Award in 2016, the Best Presentation Award from the IEICE Technical Committee on Electronics Simulation Technology in 2022, the Telecom System Technology Award from the Telecommunications Advancement Foundation in 2022, and the Best Paper Award from the IEICE Communication Society in 2007 and 2023.
\end{IEEEbiography}

\end{document}